\documentclass[superscriptaddress,twocolumn]{revtex4}
\usepackage[utf8]{inputenc}
\usepackage[T1]{fontenc}
\usepackage{graphicx}
\usepackage{amsthm}
\usepackage{amsmath}
\usepackage{mathdots}
\usepackage{caption}
\usepackage{verbatim}
\usepackage{pifont,url}
\usepackage{enumitem,amssymb}
\usepackage{multirow}
\usepackage{appendix}
\usepackage{epstopdf}
\usepackage{booktabs}
\usepackage{makecell}
\usepackage{diagbox}
\usepackage{floatrow}
\usepackage{setspace}
\usepackage{indentfirst}
\usepackage{tikz}

\tikzstyle{sNode} = [circle, fill = blue]
\tikzstyle{tNode} = [circle, fill = red]
\tikzstyle{mNode} = [circle, fill = black, scale=0.75]

\newcommand{\poneedge}[1]{
\begin{tikzpicture}[inner sep = 1.5pt, #1]
\node (1) at (0,0) [sNode]{};
\node (2) at (2,0) [tNode]{};
\path [draw, ->] (1)->(2);
\end{tikzpicture}
}

\newcommand{\ponetriangle}[1]{
\begin{tikzpicture}[inner sep = 1.5pt, #1]
\node (1) at (0,0) [sNode]{};
\node (2) at (1.5,-1.5) [mNode]{};
\node (3) at (1.5,1.5) [mNode]{};
\node (4) at (3,0) [tNode]{};
\path [draw, ->] (1)->(2);
\path [draw, ->] (1)->(3);
\path [draw, ->] (1)->(4);
\draw (2)--(4);
\draw (3)--(4);
\end{tikzpicture}
}

\newcommand{\poneindependent}[1]{
\begin{tikzpicture}[inner sep = 1.5pt, #1]
\node (1) at (0,0) [sNode]{};
\node (4) at (2,0) [tNode]{};
\node (2) at (3.5,-1.5) [mNode]{};
\node (3) at (3.5,1.5) [mNode]{};
\path [draw, ->] (1)->(4);
\draw (2)--(4);
\draw (3)--(4);
\end{tikzpicture}
}

\newcommand{\ponetriangleclique}[1]{
\begin{tikzpicture}[inner sep = 1.5pt, #1]
\node (1) at (0,0) [sNode]{};
\node (2) at (1.5,-1.5) [mNode]{};
\node (3) at (1.5,1.5) [mNode]{};
\node (4) at (3,0) [tNode]{};
\path [draw, ->] (1)->(2);
\path [draw, ->] (1)->(3);
\path [draw, ->] (1)->(4);
\draw (2)--(4);
\draw (3)--(4);
\draw (2)--(3);
\end{tikzpicture}
}

\newcommand{\poneedgeclique}[1]{
\begin{tikzpicture}[inner sep = 1.5pt, #1]
\node (1) at (0,0) [sNode]{};
\node (4) at (2,0) [tNode]{};
\node (2) at (3.5,-1.5) [mNode]{};
\node (3) at (3.5,1.5) [mNode]{};
\path [draw, ->] (1)->(4);
\draw (2)--(4);
\draw (3)--(4);
\draw (2)--(3);
\end{tikzpicture}
}

\newcommand{\ponepath}[1]{
\begin{tikzpicture}[inner sep = 1.5pt, #1]
\node (1) at (0,0) [sNode]{};
\node (2) at (2,0) [mNode]{};
\node (3) at (4,0) [tNode]{};
\path [draw, ->] (1)->(2);
\path [draw, ->] (2)->(3);
\end{tikzpicture}
}

\def\oneedge{\poneedge{scale=0.2}}

\def\onetriangle{\ponetriangle{scale=0.2}}
\def\oneindependent{\poneindependent{scale=0.2}}
\def\onepath{\ponepath{scale=0.2}}

\def\onetriangleclique{\ponetriangleclique{scale=0.2}}
\def\oneedgeclique{\poneedgeclique{scale=0.2}}

\begin{document}

\title{\textbf{Fundamental Patterns of Signal Propagation in Complex Networks}}

\author{Qitong Hu}
\affiliation{School of Mathematical Sciences, Shanghai Jiao Tong University, Shanghai, China.}
\affiliation{Ministry of Education (MOE) Funded Key Lab of Scientific and Engineering Computing, Shanghai Jiao Tong University, Shanghai, China.}
\affiliation{Shanghai Center for Applied Mathematics (SJTU Center), Shanghai Jiao Tong University, Shanghai, China.}
\author{Xiao-Dong Zhang}
\email{xiaodong@sjtu.edu.cn}
\affiliation{School of Mathematical Sciences, Shanghai Jiao Tong University, Shanghai, China.}
\affiliation{Ministry of Education (MOE) Funded Key Lab of Scientific and Engineering Computing, Shanghai Jiao Tong University, Shanghai, China.}
\affiliation{Shanghai Center for Applied Mathematics (SJTU Center), Shanghai Jiao Tong University, Shanghai, China.}

\begin{abstract}
\par
Various disasters
stem from
minor
perturbations,
such as
the spread of infectious diseases,
cascading failure
in power grids, etc.
Analyzing perturbations
is
crucial
for both theoretical and
application fields.
Previous
researchers
have
proposed
basic propagation patterns for perturbation
and
explored
the
impact of basic network motifs
on
the
collective response to
these
perturbations,
However,
the
current framework is
limited
in
its ability to
decouple
interactions,
and therefore cannot
analyze
more complex structures.
In this article,
we
establish
an effective,
robust
and powerful
propagation framework
under a general dynamic model.
This framework
reveals
common and dense network motifs
that
exert
a
critical
influence
on
signal propagation,
often spanning
orders of magnitude
compared with conclusions
generated by previous work.
Moreover,
our framework provides
a new approach to
understand
the fundamental principles
of complex systems
and
the negative feedback mechanism,
which
is of great significance for
research of
system controlling and
network resilience.
\end{abstract}

\maketitle

\section{Introduction}
\par
Signal propagation,
a
ubiquitous
phenomenon
in complex networks,
provides
theoretical
basis
for
many fields:
the spread
of infectious diseases\cite{Balcan2009MultiscaleMN},
gene regulatory dynamics\cite{Bornholdt2008BooleanNM,Balaji2006ComprehensiveAO,Rand2021GeometryOG,Karlebach2008ModellingAA,Li2014LandscapeAF},
signaling in neurodynamics\cite{Kumar2010SpikingAP}
, etc.
However,
due to
the
complexity and heterogeneity of
complex network,
decoupling
interactions
among
connections
and
dynamic models
in
signal propagation
patterns
remains conceptually difficult,
therefore
it is impossible
for
researchers
to explore
their
fundamental patterns.

\par Hens
and and colleagues\cite{Hens2019SpatiotemporalSP}
provide
a
basic
framework
for
signal propagation
based on stability theory.
They
translated
complex network topologies
into
predictions of
observed propagation patterns
and
pointed out that
the local propagation time
depends on
nodes' degree,
and
the global propagation time
relies on
the average degree of nodes
in the propagation path.
Conclusions introduced by Hens
provide
theoretical basic
to understand
some simple physical phenomena.
However,
their framework
overlooks
the 
influence
of
network motifs
on propagation patterns,
therefore
it cannot explain more complex
phenomena.

\par
\begin{figure*}[tbp]
\includegraphics[scale=0.15]{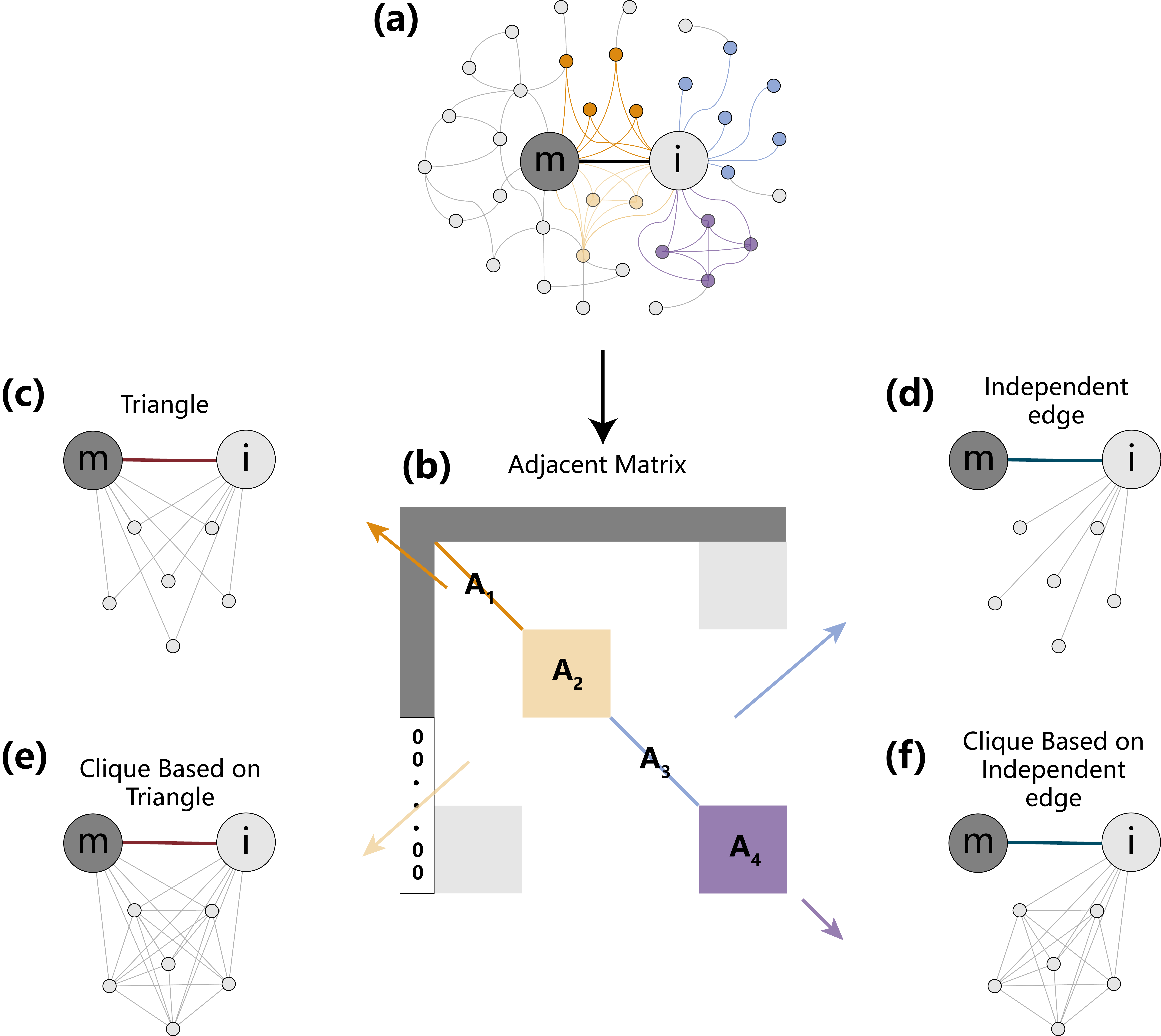}
\caption{\label{fig:1}
\textbf{Network motifs
decomposition
for
complex networks.
}
\textbf{(a)} Network with source node at $m$ and target node at $i$ having $d_i$ neighborhoods. For four colors, each color represents an motif class: triangle, independent edge, clique based on triangle and clique based on independent edge. Their specific definitions refers to (c)-(f).
\textbf{(b)} Adjacent matrix for subgraph induced by node $m$, $i$ and $i$'s neighborhood nodes. Sub-matrix $A_1$, $A_2$, $A_3$ and $A_4$ represents the connection between nodes in the same motif class. The white gray matrix represents edges between two motif class, in real network, these edges can be ignored.
\textbf{(c)} Node in triangle connects to $m$ and $i$ both, its induced graph is an independent set.
\textbf{(d)} Node in independent edge connects to $i$ only, its induced graph is an independent set.
\textbf{(e)} Node in triangle connects to $m$ and $i$ both, its induced graph is an complete graph.
\textbf{(f)} Node in independent edge connects to $i$ only, its induced graph is an complete graph.
}
\end{figure*}

Network motifs
serve as
basic building blocks
for
complex networks\cite{Milo2002NetworkMS},
and
contribute to
network complexity,
diversity,
and heterogeneity.
These
motifs
can be
categorized into
undirected
and directed motifs,
various candidate motifs
enhance
different network properties\cite{Menck2013HowBS,Menck2014HowDE,Girvan2001CommunitySI,Alon2007NetworkMT,Lambiotte2019FromNT,Battiston2021ThePO,Battiston2020NetworksBP}.
As
illustrated
in Fig.\ref{fig:1},
a complex network
can be decomposed into
four basic network motifs,
which
implies that
researching on
the signal propagation patterns
of
these
basic network motifs
can
aid in gaining
a deeper understanding
of
the essence of
signal propagation
in
complex network\cite{Lambiotte2019FromNT,Battiston2021ThePO,Battiston2020NetworksBP,StOnge2021UniversalNI}.

\par Bao and Hu\cite{Bao2022ImpactOB}
concentrated on
revealing
effect of
triangles and independent edges
in undirected motifs.
Since
triangles
constitute
a substantial proportion of
motifs
in real networks,
and greatly
influence
signal propagation patterns.
They
consider
interactions between nodes
and pointed out that
triangles and independent edges
enhance
patterns
greatly.
However, this
framework
can only
process
some
simple and sparce
network motifs,
but
is
infeasible
for more
complex
yet
common
motifs
such as clusters and cliques.

\par In this
paper,
we
introduce
an effective and powerful framework
based on
stability theory,
which 
provides
a general method
for
decoupling
interactions
in complex networks, especially for dense subgraphs.
Since we
reserve
the global topology of network,
which,
by preserving
the global network topology,
our framework
can better
elucidate
dynamic phenomena
in
real networks,
especially
these
involving
specific nodes.
Our conclusions
offer
precise definitions
for
signal propagation
and
propagation time
with regrad to
specific
network motifs,
and
can
reproduce
all the conclusions
derived
in
two previous works.
Futhermore,
following the decomposition method
illustrated in
Fig.\ref{fig:1},
we can transform
propagation analysis for
an complex real networks
into
subcases
of
basic motifs,
propagation time
of
real networks
can
be expressed by
an combination of
result for
these four cases.

\par
In contrast to
previous
theories,
our framework
is more
flexible,
interpretable,
and
controllable.
It
simplifies
the
study of
signal propagation
patterns
to
solution of
inverse matrices,
by transforming
complex
physical problems
easy
solvable
mathematical
ones.
This
greatly reduces
the difficulty of
analyzing complex structures
and
allows us to
explore
physical properties of
more complex structures.
Moreover,
our theory has successfully
elucidated
the negative feedback mechanism of
signal propagation
through
inverse laplacian transformation,
a common
but difficultly interpretable
pattern
in ecosystems.
This bring the possibilities for
investigating fundamental laws for
practical
scenarios,
such as analysis of high-order network structure, prediction of species extinction,
etc.

\section{General framework}
\subsection{General dynamic model}
\par
We consider
the
following
general
dynamic model
for $N$ nodes:
\begin{equation}
    \begin{aligned}
        \label{equ:model:1}
        \frac{dx_i(t)}{dt}=F(x_i(t))+H_1(x_i(t))\sum_{j=1}^NA_{ij}H_2(x_j(t)),
    \end{aligned}
\end{equation}
in which
$x_i$
represents
the state of node $i$, $F(x)$ characterizes the self-dynamic\cite{Castellano2007StatisticalPO,Dodds2005AGM,May1976SimpleMM,Voit2000ComputationalAO,Harush2017DynamicPO, Bao2022ImpactOB} and $H_1(x)$, $H_1(x)$ reveals interaction between two nodes\cite{Castellano2007StatisticalPO,Dodds2005AGM,May1976SimpleMM,Voit2000ComputationalAO,Harush2017DynamicPO, Bao2022ImpactOB}. Links between two nodes are expressed as a binary and symmetric matrix $A$, and degree $d_i$ is defined as $d_i=\sum_{i=1}^NA_{ij}$.
\par The unperturbed complex system is defined in a stationary state $[x_i^*]_{i=1}^N$ and satisfies following equality:
\begin{equation}
    \begin{aligned}
        \label{equ:model:2}
        0=F(x_i^*)+H_1(x_i^*)\sum_{j=1}^NA_{ij}H_2(x_j^*).
    \end{aligned}
\end{equation}
We
define
signal propagation by introducing a perturbation on the stationary state of a subset of nodes $\{i_{m_1}, i_{m_2}, \cdots, i_{m_k}\}$ with source nodes size $k$. For each initial perturbed node $i_{m_j}$, the perturbation is described by $\Delta x_{m_j}$, and shifted states is described by $x_{m_j}(t)=x_{m_j}^*+\Delta x_{m_j}$. These perturbation forces other nodes to the shifted states $x_i(t)=x_i^*+\Delta x_i(t)$, the perturbation for node $i$ satisfies following dynamic equation, which is obtained by Eq.(\ref{equ:model:1}) minus Eq.(\ref{equ:model:2}), the detailed proof has been provided in \cite{Bao2022ImpactOB,Meena2020EmergentSI}
\begin{widetext}
\begin{equation}
    \begin{aligned}
        \label{equ:model:3}
        \frac{d\Delta x_i(t)}{dt}=\left\{
        \begin{array}{l}
            0,\text{if }j\in \{i_{m_1}, i_{m_2},\cdots, i_{m_k}\},\\
            H_1(x_i^*)\left[\frac{F(x_i^*)}{H_1(x_i^*)}\right]^{'}\Delta x_i(t)+H_1(x_i^*)\sum\limits_{j=1}^NA_{ij}H_2(x_j^*)\Delta x_j(t),\text{else}.
        \end{array}
        \right.
    \end{aligned}
\end{equation}
\end{widetext}
Let $\mathbf{x}(t)=(x_1(t),\cdots,x_N(t))$. Then Eq.(\ref{equ:model:3}) can be transformed to a matrix form
\begin{equation}
    \begin{aligned}
        \label{equ:4}
        \frac{d\Delta \mathbf{x}(t)}{dt}=\tilde{\mathbf{J}}\Delta \mathbf{x}(t),
    \end{aligned}
\end{equation}
in which $\tilde{\mathbf{J}}$ is called perturbed jacobian matrix,
obtained through replacing the entries in $\{i_{m_1}, i_{m_2}, \cdots, i_{m_k}\}$-th rows of $\mathbf{J}$ by zero,
with $\mathbf{J}$ defined as follows
\begin{equation}
    \begin{aligned}
        \mathbf{J}=\text{diag}(J)+\text{diag}(S)A\text{diag}(T),
    \end{aligned}
\end{equation}
and
\begin{equation*}
    \begin{aligned}
        J=\left[H_1(x_1^*)\left[\frac{F(x_1^*)}{H_1(x_1)}\right]^{'},\right.&\cdots\left.,H_1(x_N^*)\left[\frac{F(x_N^*)}{H_1(x_N)}\right]^{'}\right],\\
        S=\left[H_1(x_1^*),\right.&\cdots\left.,H_1(x_N^*)\right],\\
        T=\left[H_2^{'}(x_1^*),\right.&\cdots\left.,H_2^{'}(x_N^*)\right].
    \end{aligned}
\end{equation*}
\subsection{Definition of Quantity and Propagation Time}
\par The solution to Eq.(\ref{equ:4}) can be expressed using an exponential function for matrix
\begin{equation}
    \begin{aligned}
        \label{equ:model:4}
        \Delta \mathbf{x}(t)=e^{\tilde{\mathbf{J}}t}\Delta \mathbf{x}(0).
    \end{aligned}
\end{equation}
Values of $e^{\tilde{\mathbf{J}}t}$ can be calculated by laplacian transformation and inverse laplacian transformation. Let $(sI-\tilde{\mathbf{J}})^{-1}$ being laplacian transformation of $e^{\tilde{\mathbf{J}}t}$, i.e.
\begin{equation*}
    \begin{aligned}
        (sI-\tilde{\mathbf{J}})^{-1}=\mathcal{L}(e^{\tilde{\mathbf{J}}t}),
    \end{aligned}
\end{equation*}
where $(sI-\tilde{\mathbf{J}})^{-1}$ can be calculated using definition of inverse matrix. The expression for $\Delta \mathbf{x}(t)$ can be obtained through the inverse laplacian transformation and Eq.(\ref{equ:model:4}) can be written as
\begin{equation}
    \begin{aligned}
        \Delta \mathbf{x}(t)=\mathcal{L}^{-1}((sI-\tilde{\mathbf{J}})^{-1})\Delta \mathbf{x}(0).
    \end{aligned}
\end{equation}
\par The matrix $\tilde{\mathbf{J}}$ is derived by deleting several rows, 
indicating that
the zero-vector is an eigenvector corresponding to the eigenvalue $0$. As $t\to \infty$, $\Delta \mathbf{x}$ will approaches to a non-zero constant vector. We define shifted state of node $i$ being the value of $\Delta x_i(t)$ when $t\to\infty$, denotes by $\Delta x_i(\infty)$, and we define the propagation time starting from perturbed nodes of node $i$ being the time when the value of $\Delta x_i(t)$ equals to $\eta-$scale of $\Delta x_i(\infty)$, denotes as $\tau_i$\cite{Hens2019SpatiotemporalSP}, i.e.
\begin{equation}
    \begin{aligned}
        \Delta x_i(\tau_i)=\eta\Delta x_i(\infty).
    \end{aligned}
\end{equation}
in which $\eta$ is a constant parameter.

\section{Dynamic models and theoretical conclusions}
\begin{table*}[tbp]
\label{con:table:1}
\centerline{
\resizebox{0.9\textwidth}{!}{
\begin{tabular}{p{1.5cm}|p{1.5cm}|p{1.5cm}|p{1.5cm}|p{2.5cm}|p{1.5cm}|p{1.5cm}|p{1.5cm}|p{1.5cm}}
\hline
\hline
    \makecell*[c]{Dynamic\\ Models}&
    \makecell*[c]{$\theta_J$}&
    \makecell*[c]{$\theta_Q$}&
    \makecell*[c]{\oneedge} &
    \makecell*[c]{\onepath} &
    \makecell*[c]{\onetriangle} &
    \makecell*[c]{\oneindependent} &
    \makecell*[c]{\onetriangleclique} &
    \makecell*[c]{\oneedgeclique}\\
\hline
    \makecell*[c]{$\mathbb{R}$}&
    \makecell*[c]{$\frac{1}{a}-1$}&
    \makecell*[c]{$-\frac{b}{a}$}&
    \makecell*[c]{$d_i^{\frac{1}{a}-1}$}&
    \makecell*[c]{$\sum\limits_{j\in P(m,i)}d_j^{\frac{1}{a}-1}$}&
    \makecell*[c]{$d_i^{\frac{1}{a}}$}&
    \makecell*[c]{$d_i^{\frac{1}{a}-1}$}&
    \makecell*[c]{$d_i^{\frac{1}{a}-1}$}&
    \makecell*[c]{$d_i^{\frac{1}{a}-1}$}\\
\hline
    \makecell*[c]{$\mathbb{P}$}&
    \makecell*[c]{$\frac{1}{a}-1$}&
    \makecell*[c]{$\frac{b}{a}$}&
    \makecell*[c]{$d_i^{\frac{1}{a}-1}$}&
    \makecell*[c]{$\sum\limits_{j\in P(m,i)}d_j^{\frac{1}{a}-1}$}&
    \makecell*[c]{$d_i^{\frac{1}{a}}$}&
    \makecell*[c]{$d_i^{\frac{1-b}{a}-1}$}&
    \makecell*[c]{$d_i^{\frac{1-b}{a}-1}$}&
    \makecell*[c]{$d_i^{\frac{1-b}{a}-1}$}\\
\hline
    \makecell*[c]{$\mathbb{E}$}&
    \makecell*[c]{$-1$}&
    \makecell*[c]{$-1$}&
    \makecell*[c]{$d_i^{-1}$}&
    \makecell*[c]{$\sum\limits_{j\in P(m,i)}d_j^{-1}$}&
    \makecell*[c]{$1$}&
    \makecell*[c]{$d_i^{-1}$}&
    \makecell*[c]{$d_i^{-1}$}&
    \makecell*[c]{$d_i^{-1}$}\\
\hline
    \makecell*[c]{$\mathbb{M}$}&
    \makecell*[c]{$-1$}&
    \makecell*[c]{$1$}&
    \makecell*[c]{$d_i^{-1}$}&
    \makecell*[c]{$\sum\limits_{j\in P(m,i)}d_j^{-1}$}&
    \makecell*[c]{$1$}&
    \makecell*[c]{$d_i^{-2}$}&
    \makecell*[c]{$d_i^{-2}$}&
    \makecell*[c]{$d_i^{-2}$}\\
\hline
    \makecell*[c]{$\mathbb{H}$}&
    \makecell*[c]{$\frac{1-b}{a}-1$}&
    \makecell*[c]{$-\frac{c}{a}$}&
    \makecell*[c]{$d_i^{\frac{1-b}{a}-1}$}&
    \makecell*[c]{$\sum\limits_{j\in P(m,i)}d_j^{\frac{1-b}{a}-1}$}&
    \makecell*[c]{$d_i^{\frac{1-b}{a}}$}&
    \makecell*[c]{$d_i^{\frac{1-b}{a}-1}$}&
    \makecell*[c]{$d_i^{\frac{1-b}{a}-1}$}&
    \makecell*[c]{$d_i^{\frac{1-b}{a}-1}$}\\
\hline
    \makecell*[c]{$\mathbb{I}$}&
    \makecell*[c]{$-1$}&
    \makecell*[c]{$\frac{1}{2}$}&
    \makecell*[c]{$d_i^{-1}$}&
    \makecell*[c]{$\sum\limits_{j\in P(m,i)}d_j^{-1}$}&
    \makecell*[c]{$1$}&
    \makecell*[c]{$d_i^{-\frac{3}{2}}$}&
    \makecell*[c]{$d_i^{-\frac{3}{2}}$}&
    \makecell*[c]{$d_i^{-\frac{3}{2}}$}\\
\hline
    \makecell*[c]{$\mathbb{B}$}&
    \makecell*[c]{$-1$}&
    \makecell*[c]{$-1$}&
    \makecell*[c]{$d_i^{-1}$}&
    \makecell*[c]{$\sum\limits_{j\in P(m,i)}d_j^{-1}$}&
    \makecell*[c]{$1$}&
    \makecell*[c]{$d_i^{-1}$}&
    \makecell*[c]{$d_i^{-1}$}&
    \makecell*[c]{$d_i^{-1}$}\\
\hline
\hline
\end{tabular}}}
\caption{\textbf{Propagation time under different dynamic models}}
\end{table*}
\subsection{Dynamic Models}
\par To estimate the quantity and propagation time of signal propagation, we will consider seven different dynamic models. These models cover fields including chemistry, biology, sociology, infectious diseases, etc.
\par \textbf{Regulatory dynamic} ($\mathbb{R}$): A dynamic model to analyze control of gene regulation, academically known as the Michaelis-Mentan dynamics. Its general formula
follows\cite{Alon2019AnIT,Karlebach2008ModellingAA,Barzel2011BinomialME}
\begin{equation}
    \begin{aligned}
        \dot{x}_i(t)=-Bx_i^a(t)+\sum_{j=1}^NA_{ij}\frac{x_j^b(t)}{1+x_j^b(t)},
    \end{aligned}
\end{equation}
in which $a$ refers to degradation when $a=1$ and dimerization when $a=2$, and $b$ is the Hill function denoting the cooperation level in gene regulatory process.
\par \textbf{Population dynamic} ($\mathbb{P}$): A dynamic model widely used in population models, biological models, etc., whose node represents the population size. Its academic name is Birth-death process and its general formula follows\cite{Novozhilov2006BiologicalAO,Hayes2004ModelingAA,Barzel2011BinomialME}
\begin{equation}
    \begin{aligned}
        \label{equ:dynamic:P}
        \dot{x}_i(t)=-Bx_i^a(t)+\sum_{j=1}^NA_{ij}x_j^b(t),
    \end{aligned}
\end{equation}
in which $b$ monitors the rate of population flow.
\par \textbf{Epidemics dynamic} ($\mathbb{E}$): A dynamic model generally used to represent the process of recovery and transportation, whose node represents the proportion of the group. Its academic name is infectious disease transmission process and its general formula is\cite{Dodds2005AGM}
\begin{equation}
    \begin{aligned}
        \dot{x}_i(t)=-Bx_i(t)+(1-x_i(t))\sum_{j=1}^NA_{ij}x_j(t),
    \end{aligned}
\end{equation}
in which $B$ is the infection rate.
\par \textbf{Mutualistic dynamic} ($\mathbb{M}$): A dynamic model to represent interactions between species on plant-pollen network, whose node represents species. The general expression for predator follows\cite{May1976SimpleMM,Holling1959SomeCO}
\begin{equation}
    \begin{aligned}
        \label{equ:dynamic:M}
        \dot{x}_i(t)=Bx_i(t)\left(1-\frac{x_i^a(t)}{C}\right)+x_i(t)\sum_{j=1}^NA_{ij}x_j(t),
    \end{aligned}
\end{equation}
in which $B$ is rate for reproducing process, $C$ reveals the opposite effect inhibiting increment of population due to limited resources and $b\le 1$.
\par \textbf{Human dynamic} ($\mathbb{H}$): A dynamic model to capture human behaviors under different environments on a macro scale, such as queuing in market or communicating on the Internet. Its general formula follows\cite{Barzel2015ConstructingMM}
\begin{equation}
    \begin{aligned}
        \dot{x}_i(t)=-Bx_i^{a+b}(t)+x_i^b(t)\sum_{j=1}^NA_{ij}(y_0-x_j^{-c}(t)),
    \end{aligned}
\end{equation}
in which $a$ and $c$ are both determined by empirical data and $b$ is an arbitrary exponent.
\par \textbf{Inhibitory dynamic} ($\mathbb{I}$): A dynamic model to represent inhibitory interactions between species on plant-pollen network, whose node represents species. The general formula for prey follows\cite{Meena2020EmergentSI}
\begin{equation}
    \begin{aligned}
        \label{equ:dynamic:I}
        \dot{x}_i(t)=-Bx_i(t)\left(1-\frac{x_i(t)}{C}\right)^2+x_i(t)\sum_{j=1}^NA_{ij}x_j(t),
    \end{aligned}
\end{equation}
$B$, $C$ and $b$ share the same physical meaning with that in Mutualistic dynamic.
\par \textbf{Biochemical dynamic} ($\mathbb{B}$): A dynamic model to characterize the concentration of protein under the consideration that heterogeneity. Its general formula follows\cite{Barzel2011BinomialME,Voit2000ComputationalAO}
\begin{equation}
    \begin{aligned}
        \dot{x}_i(t)=B-Cx_i(t)-x_i(t)\sum_{j=1}^NA_{ij}x_j(t),
    \end{aligned}
\end{equation}
in which $B$ denotes the rate describing the influx of proteins, $C$ denotes rate in correlation with
protein degradation.
\subsection{Basic Network Motifs}
\label{motif-definition}
Before
determining specific
network motifs,
we first analyze
the structure of complex networks,
focusing on
the local
propagation pattern,
i.e. the propagation process
from a node to
its neighbourhood nodes.
Following Fig.\ref{fig:1},
we can
categorize
the
neighbourhood nodes of
a target node,
denotes as $i$,
we can
classify them based on
following
two key dimensions.
The first dimension is
whether
a node
connects to
source node,
i.e an important basis for
determining whether it
forms a part of a
triangle,
while the second dimension is
whether
interactions among nodes
is dense,
i.e. an important basis
determining
whether a subgraph
constitutes
a complete graph or an independent set.
Based on
these two dimensions,
we
classify
neighbourhoods nodes
of node $i$
into four major
classes,
each representing
ones of
the four
basic network motifs,
The decomposition method
is
proposed in
Fig.\ref{fig:1},
and detailed information
about these four motifs
along with
other basic network motifs
is provided below:

\begin{enumerate}
    \item Edge(\oneedge): Information is transferred to source node's adjacent nodes.
    \item Path(\onepath): Information is transferred to the target node through intermediate nodes.
    \item Triangles(\onetriangle): Triangles are collections of nodes that are adjacent to both the source and target nodes. The influence of information on the triangle motifs affects the target node, where the generated subgraph of the triangle is an independent set.
    \item Independent edges(\oneedge): Independent edges are sets of nodes that are only adjacent to the target node and not adjacent to the source node. The influence of information on independent edges affects the target node, where the generated subgraph of independent edges is an independent set.
    \item Cliques based on triangles(\onetriangleclique): Similar to triangles, but the generated subgraph of the triangle is the complete graph.
    \item Cliques based on independent edges(\oneedgeclique): Similar to independent edges, but the generated subgraph of the independent edge is a complete graph.
\end{enumerate}
\par Next we focus on case of one source node. Following conclusions in \cite{Bao2022ImpactOB} and using $\theta_J$ and $\theta_Q$ proposed in this paper, we presents theoretical conclusion on quantity and propagation time under these seven dynamic models and list these in Table 1.

\section{Result on Basic Patterns}

\subsection{Pattern for Edge 
}
\par Message transmission along one edge is the basic pattern in signal propagation. We focus on two nodes and edge 
connecting
them, regardless of all neighbourhood nodes. This pattern allow us to monitor
signal propagation
over a short duration. Following our theoretical framework, the perturbed jacobian matrix $\tilde{\mathbf{J}}$ is
\begin{equation}
    \begin{aligned}
        \label{equ:result:edge:0}
        \tilde{\mathbf{J}}=
        \begin{bmatrix}
        0 & 0\\
        a_{21} & -b_2
        \end{bmatrix},
    \end{aligned}
\end{equation}
with
\begin{equation*}
    \begin{aligned}
        \left\{
        \begin{array}{l}
            a_{21}=H_1(x_i^*)A_{im}H_2^{'}(x_m^*),\\
            b_2=-H_1(x_i^*)\left[\frac{F(x_i^*)}{H_1(x_i^*)}\right]^{'}.
        \end{array}
        \right.
    \end{aligned}
\end{equation*}
Through definition of $\tilde{\mathbf{J}}$ in Eq.(\ref{equ:result:edge:0}), we estimate item $c_{ij}$ in $(sI-\tilde{\mathbf{J}})^{-1}$ and $c_{ij}$ satisfies following equality:
\begin{equation}
    \begin{aligned}
        \label{equ:result:edge:1}
        \left\{
        \begin{array}{l}
            c_{11}s=1,\\
            c_{11}a_{21}=c_{21}(s+b_2).
        \end{array}
        \right.
    \end{aligned}
\end{equation}
Preforming linear transformations to Eq.(\ref{equ:result:edge:1}), we can obtain exact value of $c_{21}$, which captures value of $\Delta x_i(t)$:
\begin{equation}
    \begin{aligned}
        \label{equ:result:edge:2}
        c_{21}=\frac{a_{21}}{s(s+b_2)}.
    \end{aligned}
\end{equation}
The value of $\Delta x_i(t)$ can be obtained through inverse laplacian transformation for Eq.(\ref{equ:result:edge:2}), and
\begin{equation}
    \begin{aligned}
        \label{equ:result:edge:3}
        \Delta x_i(t)&=\mathcal{L}^{-1}(c_{21})\Delta x_m\\
        &=\frac{a_{21}}{b_2}\left(1-e^{-b_2t}\right)\Delta x_m.\\
    \end{aligned}
\end{equation}
\par Shifted state for node $i$'s message is the value of Eq.(\ref{equ:result:edge:3}) when $t\to\infty$ and
\begin{equation}
    \begin{aligned}
        \Delta x_i(\infty)&=\frac{a_{21}}{b_2}\Delta x_m\\
        &=-\frac{A_{im}H_2^{'}(x_m^*)}{\left[\frac{F(x_i^*)}{H_1(x_i^*)}\right]^{'}},
    \end{aligned}
\end{equation}
while propagation time can be obtained by $\eta$-scale for Eq.(\ref{equ:result:edge:3}), i.e.
\begin{equation*}
    \begin{aligned}
        \eta&=1-e^{-b_2\tau_i},
    \end{aligned}
\end{equation*}
and
\begin{equation}
    \begin{aligned}
        \label{equ:result:edge:4}
        \tau_i&=-\frac{\ln(1-\eta)}{b_2}\\
        &=\frac{\ln(1-\eta)}{H_1(x_i^*)\left[\frac{F(x_i^*)}{H_1(x_i^*)}\right]^{'}}.
    \end{aligned}
\end{equation}
Let $\theta_J$ be the coefficient estimated by $d_i^{\theta_J}\sim \frac{1}{H_1(x_i^*)\left[\frac{F(x_i^*)}{H_1(x_i^*)}\right]^{'}}$, introduced in \cite{Bao2022ImpactOB}. Then propagation time Eq.(\ref{equ:result:edge:4}) can be simplified as
\begin{equation}
    \begin{aligned}
        \tau_i&\sim d_i^{\theta_J}.
    \end{aligned}
\end{equation}

\subsection{Pattern for Paths 
}
\par In the previous sections, we analyzed cases of local topology, which
are prominent in short-time stages. However, in signal propagation, perturbation
spread
throughout
the entire
network after a certain
period, and nodes disconnected to sources nodes are also affected by perturbations. In this section, we will reveal pattern of global topology and calculate shifted state and propagation time for basic global motifs. Paths, especially the shortest path, hold significant physical significance in network analysis. Researching signal propagation patterns for path can help us better understand the complexity of networks. Following our theoretical framework, the perturbed jacobian matrix $\tilde{\mathbf{J}}$ is
\begin{equation}
    \begin{aligned}
        \label{equ:result:path:0}
        \tilde{\mathbf{J}}=
        \begin{bmatrix}
        0 & 0 & 0\\
        a_{21} & -b_2 & 0\\
        0 & a_{32} & -b_3
        \end{bmatrix},
    \end{aligned}
\end{equation}
with
\begin{equation*}
    \begin{aligned}
        \left\{
        \begin{array}{l}
            a_{21}=H_1(x_{i_1}^*)A_{i_1m}H_2^{'}(x_m^*),\\
            a_{32}=H_1(x_i^*)A_{ii_1}H_2^{'}(x_{i_1}^*),\\
            b_2=-H_1(x_{i_1}^*)\left[\frac{F(x_{i_1}^*)}{H_1(x_{i_1}^*)}\right]^{'},\\
            b_3=-H_1(x_i^*)\left[\frac{F(x_i^*)}{H_1(x_i^*)}\right]^{'}.
        \end{array}
        \right.
    \end{aligned}
\end{equation*}
in which $i_1$ is the intermediate node. Through definition of $\tilde{\mathbf{J}}$ in Eq.(\ref{equ:result:path:0}), we estimate items $c_{ij}$ for in $(sI-\tilde{\mathbf{J}})^{-1}$ and $c_{ij}$ satisfies following equality:
\begin{equation}
    \begin{aligned}
        \label{equ:result:path:1}
        \left\{
        \begin{array}{l}
            c_{11}s=1,\\
            c_{11}a_{21}=c_{21}(s+b_2),\\
            c_{21}a_{32}=c_{31}(s+b_3).
        \end{array}
        \right.
    \end{aligned}
\end{equation}
Preforming linear transformations for Eq.(\ref{equ:result:path:1}), we can obtain value for $c_{21}$, which captures value of $\Delta x_i(t)$:
\begin{equation}
    \begin{aligned}
        \label{equ:result:path:2}
        c_{31}=\frac{a_{21}}{s(s+b_2)(s+b_3)}.
    \end{aligned}
\end{equation}
The value of $\Delta x_i(t)$ can be obtained through inverse laplacian transformation for Eq.(\ref{equ:result:path:2}), and
\begin{equation}
    \begin{aligned}
        \label{equ:result:path:3}
        \Delta x_i(t)&=\mathcal{L}^{-1}(c_{31})\Delta x_m\\
        &=\frac{a_{21}a_{32}}{b_2b_3}\left(1-\frac{b_3}{b_3-b_2}e^{-b_2t}-\frac{b_2}{b_2-b_3}e^{-b_3t}\right)\Delta x_m.
    \end{aligned}
\end{equation}
\par It can be obtained that when $b_2=b_3$, the formula of $\Delta x_i(t)$ in Eq.(\ref{equ:result:path:3}) is not available, here we provide conclusions for two limit cases. The first case we consider $b_2$, and obtain following approximation by setting $b_2\to \infty$. Shifted state for node $i$'s message is the value of Eq.(\ref{equ:result:path:3}) when $t\to\infty$ and
\begin{equation}
    \begin{aligned}
        \Delta x_i(\infty)&=\frac{a_{21}a_{32}}{b_2b_3}\Delta x_m\\
        &=\frac{A_{i_1m}A_{im_1}H_2^{'}(x_m^*)H_2^{'}(x_{i_1}^*)}{\left[\frac{F(x_{i_1}^*)}{H_1(x_{i_1}^*)}\right]^{'}\left[\frac{F(x_i^*)}{H_1(x_i^*)}\right]^{'}}\Delta x_m,
    \end{aligned}
\end{equation}
and $\Delta x_i(t)$ in Eq.(\ref{equ:result:path:3}) can be simplified as
\begin{equation}
    \begin{aligned}
        \label{equ:result:path:4}
        \Delta x_i(t)
        &=\frac{a_{21}a_{32}}{b_2b_3}\left(1-e^{-b_3t}\right)\Delta x_m,
    \end{aligned}
\end{equation}
while propagation time can be obtained by $\eta$-scale for Eq.(\ref{equ:result:path:4}), i.e.
\begin{equation*}
    \begin{aligned}
        \eta&=1-e^{-b_3\tau},
    \end{aligned}
\end{equation*}
and
\begin{equation}
    \begin{aligned}
        \label{equ:result:path:5}
        \tau&=-\frac{\ln(1-\eta)}{b_3}\\
        &=\frac{\ln(1-\eta)}{H_1(x_i^*)\left[\frac{F(x_i^*)}{H_1(x_i^*)}\right]^{'}}.
    \end{aligned}
\end{equation}
Propagation time Eq.(\ref{equ:result:path:5}) can be simplified as
\begin{equation}
    \begin{aligned}
        \tau_i&\sim d_i^{\theta_J},
    \end{aligned}
\end{equation}
The second case we consider $b_2=b_3$, there should be $c_{31}=\frac{a_{21}}{s(s+b_2)^2}$ and the previous conclusion for $\Delta x_i(t)$ is not available, value of $\Delta x_i(t)$ in Eq.(\ref{equ:result:path:3}) can be obtained through inverse laplacian transformation
\begin{equation}
    \begin{aligned}
        \label{equ:result:path:6}
        \Delta x_i(t)&=\mathcal{L}^{-1}(c_{31})\Delta x_m\\
        &=\frac{a_{21}a_{32}}{b_2^2}\left(1-(1+b_2t)e^{-b_2t}\right)\Delta x_m.
    \end{aligned}
\end{equation}
Shifted state for node $i$'s message is the value of Eq.(\ref{equ:result:path:6}) when $t\to\infty$ and
\begin{equation}
    \begin{aligned}
        \Delta x_i(\infty)&=\frac{a_{21}a_{32}}{b_2^2}\Delta x_m\\
        &=\frac{A_{i_1m}A_{im_1}H_2^{'}(x_m^*)H_2^{'}(x_{i_1}^*)}{\left[\frac{F(x_{i_1}^*)}{H_1(x_{i_1}^*)}\right]^{'}\left[\frac{F(x_i^*)}{H_1(x_i^*)}\right]^{'}}\Delta x_m,
    \end{aligned}
\end{equation}
while propagation time can be obtained by $\eta$-scale for Eq.(\ref{equ:result:path:6}), i.e.
\begin{equation*}
    \begin{aligned}
        \label{equ:result:path:7}
        \eta&=1-(1+b_2\tau)e^{-b_2\tau_i}.
    \end{aligned}
\end{equation*}
This solution is not obvious and here we provide a approximation for $\tau_i$, we let $\tau_i=(-\ln(1-\eta)+1)\frac{1}{b_2}$ and
\begin{equation*}
    \begin{aligned}
        (1+b_2\tau)e^{-b_2\tau}=\frac{2-\ln(1-\eta)}{e}(1-\eta)\approx (1-\eta).
    \end{aligned}
\end{equation*}
This means
\begin{equation}
    \begin{aligned}
        \tau_i&=(-\ln(1-\eta)+1)\frac{1}{b_2}\\
        &=\frac{-\ln(1-\eta)+1}{H_1(x_i^*)\left[\frac{F(x_i^*)}{H_1(x_i^*)}\right]^{'}}.
    \end{aligned}
\end{equation}
Propagation time Eq.(\ref{equ:result:path:7}) can be simplified as
\begin{equation}
    \begin{aligned}
        \tau_i&\sim d_{i_1}^{\theta_J}+d_i^{\theta_J}.
    \end{aligned}
\end{equation}

\subsection{Pattern for Triangles 
}
\par Triangle is a classical structure that reveals the relation between nodes and their neighbourhood nodes, it is defined by nodes that connects to both source nodes and target nodes. Triangle is also the smallest cycle in network science, and existence of cycles may lead to  collapse
of small motif,
significantly impacting
signal propagation and the stability of dynamic system. However, a simple method cannot effectively decouple the triangle structure in a network. Bao and Hu introduced a feedback framework and solved this problem under specific conditions\cite{Bao2022ImpactOB}, while our decoupling method can successfully
address this case. Following our theoretical framework, the perturbed jacobian matrix $\tilde{\mathbf{J}}$ is
\begin{equation}
    \begin{aligned}
        \label{equ:result:triangle:0}
        \tilde{\mathbf{J}}=
        \begin{bmatrix}
        0 & 0 & 0 & \cdots & 0 & 0\\
        a_{21} & -b_2 & 0 &\cdots & 0 & a_{2t}\\
        a_{31} & 0 & -b_3 &\cdots & 0 & a_{3t}\\
        \vdots & \vdots & \vdots & \ddots & \vdots & \vdots\\
        a_{t-1,1} & 0 & 0 & \cdots & -b_{t-1} & a_{t-1,t}\\
        a_{t1} & a_{t2} & a_{t3} & \cdots & a_{t,t-1} & -b_{t}
        \end{bmatrix}
    \end{aligned},
\end{equation}
in which $t=t_{im}+2$ is the index for node $i$, $i_1,i_2,i_{t-1}$ are indices of triangles for node $i$, $t_{im}$ is number of triangles for node $i$ and $m$, and
\begin{equation*}
    \begin{aligned}
        \left\{
        \begin{array}{l}
            a_{k1}=H_1(x_{i_k}^*)A_{im}H_2^{'}(x_m^*),\\
            a_{tk}=H_1(x_i^*)A_{ti_k}H_2^{'}(x_{i_k}^*),\\
            a_{kt}=H_1(x_{i_k}^*)A_{i_ki}H_2^{'}(x_i^*),\\
            b_k=-H_1(x_{i_k}^*)\left[\frac{F(x_{i_k}^*)}{H_1(x_{i_k}^*)}\right]^{'},\\
            b_t=-H_1(x_{i}^*)\left[\frac{F(x_{i}^*)}{H_1(x_{i}^*)}\right]^{'}.
        \end{array}
        \right.
        ,k=2,t_{im}+1.
    \end{aligned}
\end{equation*}
Through deinfition of $\tilde{\mathbf{J}}$ in Eq.(\ref{equ:result:triangle:0}), we estiate items $c_{ij}$ for in $(sI-\tilde{\mathbf{J}})^{-1}$ and $c_{ij}$ satisfies following equality:
\begin{equation}
    \begin{aligned}
        \label{equ:result:triangle:1}
        \left\{
        \begin{array}{l}
            c_{11}s=1,\\
            c_{11}a_{k1}+c_{t1}a_{kt}=c_{k1}(s+b_k),k=2,t_{im}+1,\\
            a_{t1}c_{11}+\sum\limits_{k=2}^{t_{im}+1}a_{tk}c_{k1}=c_{t1}(s+b_t).
        \end{array}
        \right.
    \end{aligned}
\end{equation}
Preforming linear transformations to Eq.(\ref{equ:result:triangle:1}), we can obtain value for $c_{21}$, which captures value of $\Delta x_i(t)$:
\begin{equation}
    \begin{aligned}
        \label{equ:result:triangle:2}
        c_{t1}&=\frac{\frac{a_{t1}}{s}+\sum\limits_{k=2}^{t_{im}+1}\frac{a_{tk}a_{k1}}{s(s+b_k)}}{s+b_t-\sum\limits_{k=2}^{t_{im}+1}\frac{a_{tk}a_{kt}}{s+b_k}}\\
        &\approx \frac{a_{t1}}{s(s+\tilde{b}_t)}+\sum\limits_{k=2}^{t_{im}+1}\frac{a_{tk}a_{k1}}{s(s+b_k)(s+\tilde{b}_t)},
    \end{aligned}
\end{equation}
here we ignore $s$ in the denominator under the condition $b_k$ is large, i.e. $\theta_J<0$, and
\begin{equation*}
    \begin{aligned}
        \tilde{b}_t&=b_t-\sum\limits_{k=2}^{t_{im}+1}\frac{a_{tk}a_{kt}}{b_k}\\
        &=-H_1(x_{i}^*)\left[\frac{F(x_{i}^*)}{H_1(x_{i}^*)}\right]^{'}\left(1-\sum\limits_{k=2}^{t_{im}+1}A_{i_ki}A_{ii_k}C(x_i^*)C(x_{i_k}^*)\right),
    \end{aligned}
\end{equation*}
in which $C(x)=\frac{H_2^{'}(x)}{\left[\frac{F(x)}{H_1(x)}\right]^{'}}$. The value of $\Delta x_i(t)$ can be approximately obtained through inverse laplacian transformation for Eq.(\ref{equ:result:triangle:2}), and
\begin{equation}
    \begin{aligned}
        \label{equ:result:triangle:3}
        \Delta x_i(t)&=\mathcal{L}^{-1}(c_{t1})\Delta x_m\\
        &\approx \frac{a_{t1}}{\tilde{b}_t}\left(1-e^{-\tilde{b}_tt}\right)\\
        &\quad+\sum\limits_{k=2}^{t_{im}+1}\frac{a_{tk}a_{k1}}{b_k\tilde{b}_t}\left(1-\frac{b_k}{b_k-\tilde{b}_t}e^{-\tilde{b}_tt}-\frac{\tilde{b}_t}{\tilde{b}_t-b_k}e^{-\tilde{b}_kt}\right).
    \end{aligned}
\end{equation}
\par Shifted state for node $i$'s message is the value of Eq.(\ref{equ:result:triangle:3}) when $t\to\infty$ and
\begin{equation}
    \begin{aligned}
        \Delta x_i(\infty)&=\frac{a_{t1}}{\tilde{b}_t}+\sum\limits_{k=2}^{t_{im}+1}\frac{a_{tk}a_{k1}}{b_k\tilde{b}_t}\\
        &=-\frac{A_{im}H_2^{'}(x_m^*)}{\left[\frac{F(x_{i}^*)}{H_1(x_{i}^*)}\right]^{'}}\frac{1+\sum\limits_{k=2}^{t_{im}+1}\frac{A_{ii_k}A_{i_km}}{A_{im}}C(x_{i_k}^*)}{1-\sum\limits_{k=2}^{t_{im}+1}A_{i_ki}A_{ii_k}C(x_i^*)C(x_{i_k}^*)},
    \end{aligned}
\end{equation}
propagation time can be obtained by $\eta$-scale for Eq.(\ref{equ:result:triangle:3}), however, for triangles this value is hard to be approximated, and here we provide conclusions for two limit case. The first case we consider $b_k$ is large, then we can obtain following approximation of Eq.(\ref{equ:result:triangle:3}) by setting $b_k\to \infty$
\begin{equation}
    \begin{aligned}
        \label{equ:result:triangle:4}
        \Delta x_i(t)
        &\approx \frac{a_{t1}}{\tilde{b}_t}\left(1-e^{-\tilde{b}_tt}\right)+\sum\limits_{k=2}^{t_{im}+1}\frac{a_{tk}a_{k1}}{b_k\tilde{b}_t}\left(1-e^{-\tilde{b}_tt}\right)\\
        &=\frac{a_{t1}}{\tilde{b}_t}\left(1+\frac{a_{tk}a_{k1}}{a_{t1}b_k}\right)\left(1-e^{-\tilde{b}_tt}\right),
    \end{aligned}
\end{equation}
while propagation time can be obtained by $\eta$-scale for Eq.(\ref{equ:result:triangle:4}), i.e.
\begin{equation*}
    \begin{aligned}
        \eta&=\left(1-e^{-\tilde{b}_t\tau_i}\right),
    \end{aligned}
\end{equation*}
and
\begin{equation}
    \begin{aligned}
        \label{equ:result:triangle:5}
        \tau_i&=-\frac{\ln(1-\eta)}{\tilde{b}_t}\\
        &=\frac{\ln(1-\eta)}{H_1(x_{i}^*)\left[\frac{F(x_{i}^*)}{H_1(x_{i}^*)}\right]^{'}}\frac{1}{1-\sum\limits_{k=2}^{t_{im}+1}A_{i_ki}A_{ii_k}C(x_i^*)C(x_{i_k}^*)}.
    \end{aligned}
\end{equation}
Let $\theta_Q$ be the coefficient estimated by $d_i^{\theta_Q-1}\sim C(x_i^*)$, which is introduced in \cite{Bao2022ImpactOB}, and for $\theta_Q<0$ propagation time Eq.(\ref{equ:result:triangle:5}) can be simplified as
\begin{equation}
    \begin{aligned}
        \tau_i&\sim d_i^{\theta_J},
    \end{aligned}
\end{equation}
for $\theta_Q>0$ propagation time Eq.(\ref{equ:result:triangle:5}) can be simplified as
\begin{equation}
    \begin{aligned}
        \tau_i&\sim d_i^{\theta_J-\theta_Q}.
    \end{aligned}
\end{equation}
The second case we consider $b_t$ is large, then we can obtain following approximation by setting $b_t\to \infty$. This case differs from the first case, since each $b_k$ shares different values and we cannot summary them together simply. Here we introduce a method named linear near, for random variables $[x_i]_{i=1}^N$ and a nonlinear function $f(x)$, if $f^{''}(x)$ approaches to $0$ in range of $x_i$, we can judge $f(x)$ as a linear function, this claim can be easily check through Taylor expansion for $f(x)$ near $\frac{1}{N}\sum\limits_{i=1}^Nf(x_i^*)$. Using linear near method, we can approximate $\Delta x_i(t)$ in Eq.(
\ref{equ:result:triangle:3})
to an exponential type function as follows
\begin{widetext}
\begin{equation*}
    \begin{aligned}
        \label{equ:result:triangle:6}
        \Delta x_i(t)
        &\approx \frac{a_{t1}}{\tilde{b}_t}\left(1-e^{-\tilde{b}_tt}\right)+\sum\limits_{k=2}^{t_{im}+1}\frac{a_{tk}a_{k1}}{b_k\tilde{b}_t}\left(1-e^{-b_kt}\right)\\
        &=\frac{a_{t1}}{\tilde{b}_t}\left(1+\sum\limits_{k=2}^{t_{im}+1}\frac{a_{tk}a_{k1}}{a_{t1}b_k}\right)\left(1-\frac{e^{-\tilde{b}_tt}}{1+\sum\limits_{k=2}^{t_{im}+1}\frac{a_{tk}a_{k1}}{a_{t1}b_k}}-\sum\limits_{k=2}^{t_{im}+1}\frac{\frac{a_{tk}a_{k1}}{a_{t1}b_k}e^{-b_kt}}{1+\sum\limits_{k=2}^{t_{im}+1}\frac{a_{tk}a_{k1}}{a_{t1}b_k}}\right)\\
        &\approx\frac{a_{t1}}{\tilde{b}_t}\left(1+\sum\limits_{k=2}^{t_{im}+1}\frac{a_{tk}a_{k1}}{a_{t1}b_k}\right)\left(1-\exp\left(-{\left(\tilde{b}_t+\sum\limits_{k=2}^{t_{im}+1}\frac{a_{tk}a_{k1}}{a_{t1}b_k}b_k\right)}\Big/{\left(1+\sum\limits_{k=2}^{t_{im}+1}\frac{a_{tk}a_{k1}}{a_{t1}b_k}\right)}t\right)\right),
    \end{aligned}
\end{equation*}
\end{widetext}
in which $\sum\limits_{k=2}^{t_{im}+1}\frac{A_{ii_k}A_{i_km}}{A_{im}}C(x_{i_k}^*)b_k$ is much smaller than $b_t$ since we consider $b_t$ is very large. Propagation time can be obtained by $\eta$-scale for Eq.(\ref{equ:result:triangle:6}), i.e.
\begin{equation*}
    \begin{aligned}
        \eta&=1-\exp\left(-\frac{\tilde{b}_t}{1+\sum\limits_{k=2}^{t_{im}+1}\frac{a_{tk}a_{k1}}{a_{t1}b_k}}\tau_i\right),
    \end{aligned}
\end{equation*}
and
\begin{equation}
    \begin{aligned}
        \label{equ:result:triangle:7}
        \tau_i&=-\ln(1-\eta)\left(\frac{a_{t1}}{\tilde{b}_t}+\sum\limits_{k=2}^{t_{im}+1}\frac{a_{tk}a_{k1}}{b_k\tilde{b}_t}\right)\\
        &=\frac{\ln(1-\eta)}{H_1(x_i^*)\left[\frac{F(x_i^*)}{H_1(x_i^*)}\right]^{'}}\frac{1+\sum\limits_{k=2}^{t_{im}+1}\frac{A_{ii_k}A_{i_km}}{A_{im}}C(x_{i_k}^*)}{1-\sum\limits_{k=2}^{t_{im}+1}A_{i_ki}A_{ii_k}C(x_i^*)C(x_{i_k}^*)}.
    \end{aligned}
\end{equation}
For $\theta_J<0$ and $\theta_Q<0$, propagation time Eq.(\ref{equ:result:triangle:7}) can be simplified as
\begin{equation}
    \begin{aligned}
        \label{equ:result:triangle:8}
        \tau_i&\sim d_i^{\theta_J+1},
    \end{aligned}
\end{equation}
and for $\theta_J<0$ and $\theta_Q>0$, propagation time Eq.(\ref{equ:result:triangle:7}) can be simplified as
\begin{equation}
    \begin{aligned}
        \tau_i&\sim d_i^{\theta_J-\theta_Q+1}.
    \end{aligned}
\end{equation}
\par Effect of triangles have been identified by Bao and Hu, in Fig.2(g) of main text\cite{Bao2022ImpactOB}.

\subsection{Pattern for Independent edges 
}
\par Definition of independent edges is related to that of triangles, independent edge refers to nodes connected to target nodes but disconnected from sources nodes. Bao and Hu's theory
highlights the existence of message feedback,
which imlplies that message is undirected and maybe enlarged on only one edge. Following our theoretical framework, the perturbed jacobian matrix $\tilde{\mathbf{J}}$ is
\begin{equation}
    \begin{aligned}
        \label{equ:result:independent:0}
        \tilde{\mathbf{J}}=
        \begin{bmatrix}
        0 & 0 & 0 & \cdots & 0 & 0\\
        0 & -b_2 & 0 &\cdots & 0 &a_{2t}\\
        0 & 0 & -b_3 &\cdots & 0 &a_{3t}\\
        \vdots & \vdots & \vdots & \ddots & \vdots & \vdots\\
        0 & 0 & 0 &\cdots & -b_{t-1} &a_{t-1,t}\\
        a_{t1} & a_{t2} & a_{t3} & \cdots & a_{t,t-1} & -b_{t}
        \end{bmatrix}
    \end{aligned},
\end{equation}
in which $t=s_{im}+2$ is the index for node $i$, $i_1,i_2,i_{t-1}$ are indexs of independent edges for node $i$, $s_{im}$ is number of independent edges for node $i$ and $m$, and
\begin{equation*}
    \begin{aligned}
        \left\{
        \begin{array}{l}
            a_{k1}=H_1(x_{i_k}^*)A_{im}H_2^{'}(x_m^*),\\
            a_{tk}=H_1(x_i^*)A_{ii_k}H_2^{'}(x_{i_k}^*),\\
            a_{kt}=H_1(x_{i_k}^*)A_{i_ki}H_2^{'}(x_i^*),\\
            b_k=-H_1(x_{i_k}^*)\left[\frac{F(x_{i_k}^*)}{H_1(x_{i_k}^*)}\right]^{'},\\
            b_t=-H_1(x_{i}^*)\left[\frac{F(x_{i}^*)}{H_1(x_{i}^*)}\right]^{'}.
        \end{array}
        \right.
        ,k=2,s_{im}+1.
    \end{aligned}
\end{equation*}
Through definition of $\tilde{\mathbf{J}}$ in Eq.(\ref{equ:result:independent:0}), we calculate items $c_{ij}$ for in $(sI-\tilde{\mathbf{J}})^{-1}$ and $c_{ij}$ satisfies following equality:
\begin{equation}
    \begin{aligned}
        \label{equ:result:independent:1}
        \left\{
        \begin{array}{l}
            c_{11}s=1,\\
            c_{t1}a_{kt}=c_{k1}(s+b_k),k=2,t_{im}+1,\\
            a_{t1}c_{11}+\sum\limits_{k=2}^{s_{im}+1}a_{tk}c_{k1}=c_{t1}(s+b_t).
        \end{array}
        \right.
    \end{aligned}
\end{equation}
Preforming linear transformations for Eq.(\ref{equ:result:independent:1}), we can obtain value for $c_{21}$, which captures value of $\Delta x_i(t)$:
\begin{equation*}
    \begin{aligned}
        \label{equ:result:independent:2}
        c_{t1}&=\frac{a_{t1}}{s\left(s+b_t-\sum\limits_{k=2}^{s_{im}+1}\frac{a_{tk}a_{kt}}{s+b_k}\right)}\\
        &\approx \frac{a_{t1}}{s\left(s+\tilde{b}_t\right)},
    \end{aligned}
\end{equation*}
here we ignore $s$ in the denominator under the condition $b_k$ is large, i.e. $\theta_J<0$, and
\begin{equation*}
    \begin{aligned}
        \tilde{b}_t&=b_t-\sum\limits_{k=2}^{s_{im}+1}\frac{a_{tk}a_{kt}}{b_k}\\
        &=-H_1(x_{i}^*)\left[\frac{F(x_{i}^*)}{H_1(x_{i}^*)}\right]^{'}\left(1-\sum\limits_{k=2}^{s_{im}+1}A_{i_ki}A_{ii_k}C(x_i^*)C(x_{i_k}^*)\right).
    \end{aligned}
\end{equation*}
The value of $\Delta x_i(t)$ can be approximately obtained through inverse laplacian transformation for Eq.(\ref{equ:result:independent:2}), and
\begin{equation}
    \begin{aligned}
        \label{equ:result:independent:3}
        \Delta x_i(t)&=\mathcal{L}^{-1}(c_{t1})\Delta x_m\\
        &=\frac{a_{t1}}{\tilde{b}_t}e^{-\tilde{b}_tt}.
    \end{aligned}
\end{equation}
\par Shifted state for node $i$'s message is the value of Eq.(\ref{equ:result:independent:3}) when $t\to\infty$ and
\begin{equation}
    \begin{aligned}
        \Delta x_i(\infty)
        &=\frac{a_{t1}}{\tilde{b}_t}\\
        &=-\frac{A_{ii_k}H_2^{'}(x_{i_k}^*)}{\left[\frac{F(x_{i}^*)}{H_1(x_{i}^*)}\right]^{'}}\frac{1}{1-\sum\limits_{k=2}^{s_{im}+1}A_{i_ki}A_{ii_k}C(x_i^*)C(x_{i_k}^*)},
    \end{aligned}
\end{equation}
while propagation time can be obtained by $\eta$-scale for Eq.(\ref{equ:result:independent:3}), i.e.
\begin{equation*}
    \begin{aligned}
        \eta&=1-e^{-\tilde{b}_t\tau_i},
    \end{aligned}
\end{equation*}
and
\begin{equation}
    \begin{aligned}
        \label{equ:result:independent:4}
        \tau_i&=-\frac{\ln(1-\eta)}{\tilde{b}_t}\\
        &=\frac{\ln(1-\eta)}{H_1(x_{i}^*)\left[\frac{F(x_{i}^*)}{H_1(x_{i}^*)}\right]^{'}}\frac{1}{1-\sum\limits_{k=2}^{s_{im}+1}A_{i_ki}A_{ii_k}C(x_{i_k}^*)C(x_i^*)}.
    \end{aligned}
\end{equation}
For $\theta_J<0$ and $\theta_Q<0$ propagation time Eq.(\ref{equ:result:independent:4}) can be simplified as
\begin{equation}
    \begin{aligned}
        \tau_i&\sim d_i^{\theta_J},
    \end{aligned}
\end{equation}
and for $\theta_J<0$ and $\theta_Q>0$ propagation time Eq.(\ref{equ:result:independent:4}) can be simplified as
\begin{equation}
    \begin{aligned}
        \tau_i&\sim d_i^{\theta_J-\theta_Q}.
    \end{aligned}
\end{equation}
\par Effect of triangles have been identified by Bao and Hu, in Fig.2(e) of main text\cite{Bao2022ImpactOB}.

\subsection{Pattern for Cliques Based on Triangle 
}
\begin{figure*}[tbp]
\includegraphics[scale=0.33]{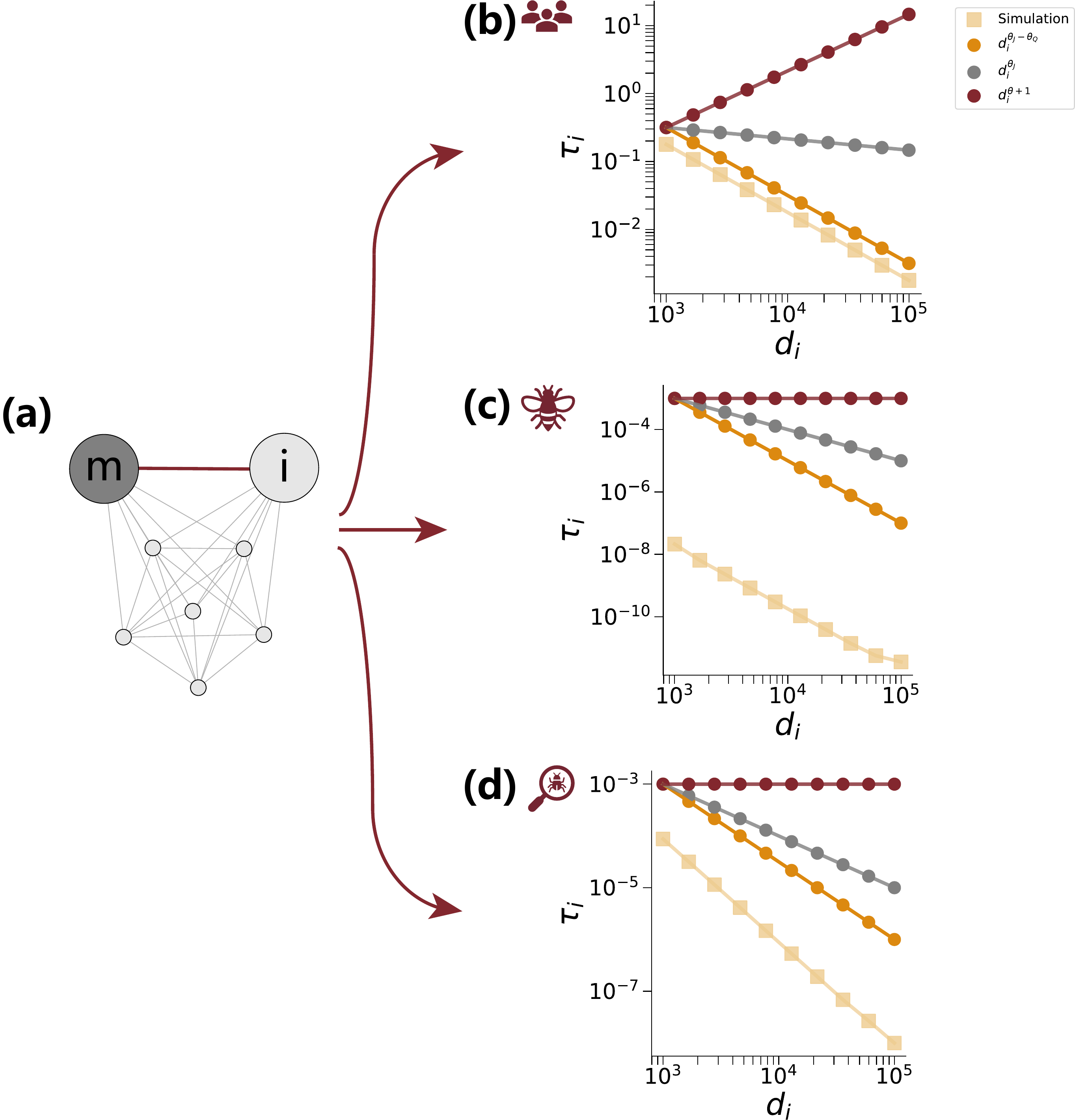}
\caption{
\label{fig:2}
\textbf{Local propagation time for cliques based on triangles.}
\textbf{(a)} Network with perturbation at $m$ and target to node i having $d_i$ edges, each node connecting to $m$ and $i$ connects to each other.
\textbf{(b)} Propagation time $\tau_i$ simulated on network (a) and population dynamics proposed proposed in Eq.(\ref{equ:dynamic:P}): $\dot{x}_i(t)=-Bx_i^a(t)+\sum\limits_{j=1}^NA_{ij}x_j^b(t)$,  with $B=0.1, a=1.2, b=1, \alpha=0.01(\theta_J, \theta_Q<0)$. Comparison of scaling coefficient Eqs.(\ref{equ:result:triangle:8}), (\ref{equ:hens:1}), (\ref{equ:ours:1}) and simulation. The trend of simulation propagation time agrees to our prediction $\tau_i\sim d_i^{\theta_J-\theta_Q}$ in Eq.(\ref{equ:ours:1}), but disagrees the prediction $\tau_i\sim d_i^{\theta_J}$(gray) in Eq.(\ref{equ:hens:1}) and $\tau_i\sim d_i^{\theta_J+1}$(brown) in Eq.(\ref{equ:result:triangle:8}).
\textbf{(c)} Simulation on network (a) and Mutualistic dynamics proposed in Eq.(\ref{equ:dynamic:M}): $\dot{x}_i(t)=-Bx_i(t)\left(1-\frac{x_i(t)}{C}\right)+x_i(t)\sum\limits_{j=1}^NA_{ij}x_j(t)$,  with $B=0.1, C=2, \alpha=0.01(\theta_J, \theta_Q<0)$.
\textbf{(d)} Simulation on network (a) and inhibitory dynamics proposed in Eq.(\ref{equ:dynamic:I}): $\dot{x}_i(t)=-Bx_i(t)\left(1-\frac{x_i(t)}{C}\right)^2+x_i(t)\sum\limits_{j=1}^NA_{ij}x_j(t)$,  with $B=0.1, C=2, \alpha=0.01(\theta_J, \theta_Q<0)$.
}
\end{figure*}
\par In some theoretical network models, such as Erd\H{o}s-R\'{e}nyi model, Barab\'{a}si-Albert model, Chung-Lu model, etc.,  rare cliques structure exist, this means that probabilistic methods are not available for the analysis of cliques. However, community structure typically occur in the real world, and in a strict definition, community and clique share the same physical meaning. The theory of cliques can be directly applied to analysis of community structures. Our framework can effectively handle this case and decouple interactions within clique.
\par First we consider the case of cliques based on triangles, this case is common in real world since $k-$core determines stability of certain dynamic systems, and $k-$core can be viewed as an clique. Following our theoretical framework, the perturbed jacobian matrix $\tilde{\mathbf{J}}$ is
\begin{equation}
    \begin{aligned}
        \label{equ:result:cliquetri:0}
        \tilde{\mathbf{J}}=
        \begin{bmatrix}
        0 & 0 & 0 & \cdots & 0 & 0\\
        a_{21} & -b_2 & a_{23} &\cdots & a_{2,t-1} & a_{2t}\\
        a_{31} & a_{32} & b_3 &\cdots & a_{3,t-11} & a_{3t}\\
        \vdots & \vdots & \vdots & \ddots & \vdots & \vdots\\
        a_{t-1,1} & a_{t-1,2} & a_{t-1,3} & \cdots & -b_{t-1} & -a_{t-1,t}\\
        a_{t1} & a_{t2} & a_{t3} & \cdots & a_{t,t-1} & -b_{t}
        \end{bmatrix}
    \end{aligned},
\end{equation}
in which $t=t_{im}+2$ is the index for node $i$, $i_1,i_2,i_{t-1}$ are indexs of triangles for node $i$, and
\begin{equation*}
    \begin{aligned}
        \left\{
        \begin{array}{l}
            a_{t1}=H_1(x_i^*)A_{im}H_2^{'}(x_m^*),\\
            a_{k1}=H_1(x_{i_k}^*)A_{im}H_2^{'}(x_m^*),\\
            a_{tk}=H_1(x_i^*)A_{ii_k}H_2^{'}(x_{i_k}^*),\\
            a_{kt}=H_1(x_{i_k}^*)A_{i_ki}H_2^{'}(x_i^*),\\
            b_k=-H_1(x_{i_k}^*)\left[\frac{F(x_{i_k}^*)}{H_1(x_{i_k}^*)}\right]^{'},\\
            b_t=-H_1(x_{i}^*)\left[\frac{F(x_{i}^*)}{H_1(x_{i}^*)}\right]^{'}.
        \end{array}
        \right.
        ,k=2,t_{im}+1.
    \end{aligned}
\end{equation*}
Through definition of $\tilde{\mathbf{J}}$ in Eq.(\ref{equ:result:cliquetri:0}), we calculate items $c_{ij}$ for in $(sI-\tilde{\mathbf{J}})^{-1}$ and $c_{ij}$ satisfies following linear system of equation:
\begin{equation}
    \begin{aligned}
        \label{equ:result:cliquetri:1}
        \begin{bmatrix}
        -a_{21} & s+b_2 &\cdots & -a_{2,t-1} & -a_{2t}\\
        -a_{31} & -a_{32} &\cdots & -a_{3,t-11} & -a_{3t}\\
        \vdots & \vdots & \ddots & \vdots & \vdots\\
        -a_{t-1,1} & -a_{t-1,2} & \cdots & s+b_{t-1} & -a_{t-1,t}\\
        -a_{t1} & -a_{t2} & \cdots & -a_{t,t-1} & s+b_{t}
        \end{bmatrix}
        \begin{bmatrix}
        \frac{1}{s} \\
        c_{21} \\
        \vdots \\
        c_{t-1,1} \\
        c_{t1}
        \end{bmatrix}=0.
    \end{aligned}
\end{equation}
This linear system Eq.(\ref{equ:result:cliquetri:1}) can be simplified after multiplying the following matrix to the left
\begin{equation*}
    \begin{aligned}
        \begin{bmatrix}
            1 & 0 &\cdots & 0 & -f_2\\
            0 & 1 &\cdots & 0 & -f_3\\
            \vdots & \vdots & \ddots & \vdots & \vdots\\
            0 & 0 & \cdots & 1 & -f_{t-1}\\
            0 & 0 & \cdots & 0 & 1
        \end{bmatrix},
    \end{aligned}
\end{equation*}
in which $f_k=-\frac{H_1(x_{i_k}^*)}{H_1(x_i^*)},k=2,t_{im}+1$, we obtain
\begin{equation}
    \begin{aligned}
        \label{equ:result:cliquetri:2}
        \begin{bmatrix}
        0 & s+b_2+f_2a_{t2} &\cdots & -a_{2t}-f_2(s+b_t)\\
        0 & 0 &\cdots & -a_{3t}-f_3(s+b_t)\\
        \vdots & \vdots & \ddots & \vdots\\
        0 & 0& \cdots & -a_{t-1,t}-f_{t-1}(s+b_t)\\
        -a_{t1} & -a_{t2} & \cdots & s+b_{t}
        \end{bmatrix}
        \begin{bmatrix}
        \frac{1}{s} \\
        c_{21} \\
        \vdots \\
        c_{t1}
        \end{bmatrix}=0.
    \end{aligned}
\end{equation}
Preforming linear transformations for Eq.(\ref{equ:result:cliquetri:2}), we can obtain value for $c_{21}$, which captures value of $\Delta x_i(t)$:
\begin{equation}
    \begin{aligned}
        \label{equ:result:cliquetri:3}
        c_{t1}&=\frac{a_{t1}\frac{1}{s}}{s+b_t-\sum\limits_{k=2}^{t_{im}+1}\frac{a_{tk}\left(a_{kt}+f_k(s+b_t)\right)}{s+b_k+f_ka_{tk}}}\\
        &=\frac{a_{t1}\frac{1}{s}}{s\left(1-\sum\limits_{k=2}^{t_{im}+1}\frac{a_{tk}f_k}{s+b_k+f_ka_{tk}}\right)+b_t-\sum\limits_{k=2}^{t_{im}+1}\frac{a_{tk}a_{kt}+f_kb_t}{s+b_k+f_ka_{tk}}}\\
        &\approx \frac{\frac{a_{t1}}{1-e_1}}{s\left(s+\frac{b_t-e_2}{1-e_1}\right)},
    \end{aligned}
\end{equation}
where we ignore $s$ in the denominator under the condition $b_k+d_ka_{tk}$ is large, i.e.
\begin{equation*}
    \begin{aligned}
        b_k+f_ka_{tk}&=-H_1(x_{i_k}^*)\left[\frac{F(x_{i_k}^*)}{H_1(x_{i_k}^*)}\right]^{'}-H_1(x_{i_k}^*)A_{ti_k}H_2^{'}(x_{i_k}^*)\\
        &=-H_1(x_{i_k}^*)\left[\frac{F(x_{i_k}^*)}{H_1(x_{i_k}^*)}\right]^{'}\left(1+A_{ii_k}C(x_{i_k}^*)\right),
    \end{aligned}
\end{equation*}
this needs $\theta_J<0$, and
\begin{equation*}
    \begin{aligned}
        e_1&=\sum\limits_{k=2}^{t_{im}+1}\frac{a_{tk}f_k}{b_k+f_ka_{tk}}\\
        &=\sum\limits_{k=2}^{t_{im}+1}\frac{A_{ii_k}}{C^{-1}(x_{i_k}^*)+A_{ii_k}},
    \end{aligned}
\end{equation*}
and
\begin{equation*}
    \begin{aligned}
        e_2&=\sum\limits_{k=2}^{t_{im}+1}\frac{a_{tk}a_{kt}+a_{tk}f_kb_t}{b_k+f_ka_{tk}}\\
        &=-H_1(x_i^*)H_2^{'}(x_i^*)\sum\limits_{k=2}^{t_{im}+1}\frac{A_{ii_k}A_{i_ki}+A_{ii_k}C^{-1}(x_{i}^*)}{C^{-1}(x_{i_k}^*)+A_{ii_k}},
    \end{aligned}
\end{equation*}
The value of $\Delta x_i(t)$ can be approximately obtained through inverse laplacian transformation for Eq.(\ref{equ:result:cliquetri:3}), and
\begin{equation}
    \begin{aligned}
        \label{equ:result:cliquetri:4}
        \Delta x_i(t)&=\mathcal{L}^{-1}(c_{t1})\Delta x_m\\
        &=\frac{\frac{a_{t1}}{1-e_1}\left(1-\exp\left(-\left(b_t-\frac{e_2}{1-e_1}\right)t\right)\right)}{\frac{b_t-e_2}{1-e_1}}\Delta x_m.
    \end{aligned}
\end{equation}
\par Shifted state for node $i$'s message is the value of Eq.(\ref{equ:result:cliquetri:4}) when $t\to\infty$ and
\begin{equation}
    \begin{aligned}
        \Delta x_i(\infty)&=\frac{a_{t1}}{b_t-e_2}\Delta x_m\\
        &=\frac{A_{im}H_2^{'}(x_m^*)\Delta x_m}{-\left[\frac{F(x_{i}^*)}{H_1(x_{i}^*)}\right]^{'}+H_2^{'}(x_i^*)\sum\limits_{k=2}^{t_{im}+1}\frac{A_{ii_k}A_{i_ki}+A_{ii_k}C^{-1}(x_{i}^*)}{C^{-1}(x_{i_k}^*)+A_{ii_k}}},
    \end{aligned}
\end{equation}
while propagation time can be obtained by $\eta$-scale for Eq.(\ref{equ:result:cliquetri:4}), i.e.
\begin{equation*}
    \begin{aligned}
        \eta&=1-\exp\left(-\left(\frac{b_t-e_2}{1-e_1}\right)\tau_i\right),
    \end{aligned}
\end{equation*}
and
\begin{equation}
    \begin{aligned}
        \label{equ:result:cliquetri:5}
        \tau_i&=-\frac{\ln(1-\eta)}{\frac{b_t-e_2}{1-e_1}}\\
        &=\frac{\ln(1-\eta)}{H_1(x_{i}^*)\left[\frac{F(x_{i}^*)}{H_1(x_{i}^*)}\right]^{'}}\frac{1-\sum\limits_{k=2}^{t_{im}+1}\frac{A_{ii_k}}{C^{-1}(x_{i_k}^*)+A_{ii_k}}}{1+C(x_i^*)\sum\limits_{k=2}^{t_{im}+1}\frac{A_{ii_k}A_{i_ki}+A_{ii_k}C^{-1}(x_{i}^*)}{C^{-1}(x_{i_k}^*)+A_{ii_k}}}.
    \end{aligned}
\end{equation}
For $\theta_J<0$ and $\theta_Q<0$ propagation time Eq.(\ref{equ:result:cliquetri:5}) can be simplified as
\begin{equation}
    \begin{aligned}
        \label{equ:hens:1}
        \tau_i&\sim d_i^{\theta_J},
    \end{aligned}
\end{equation}
and for $\theta_J<0$ and $\theta_Q>0$ propagation time Eq.(\ref{equ:result:cliquetri:5}) can be simplified as
\begin{equation}
    \begin{aligned}
        \label{equ:ours:1}
        \tau_i&\sim d_i^{\theta_J-\theta_Q}.
    \end{aligned}
\end{equation}
\par To identity influence of clique based on triangles, we use the similar methods proposed by Bao and Hu and design such a simulation, the result has been shown in Fig.\ref{fig:2}.

\subsection{Pattern for Cliques Based on Independent Edge
}
\begin{figure*}[htbp]
\includegraphics[scale=0.33]{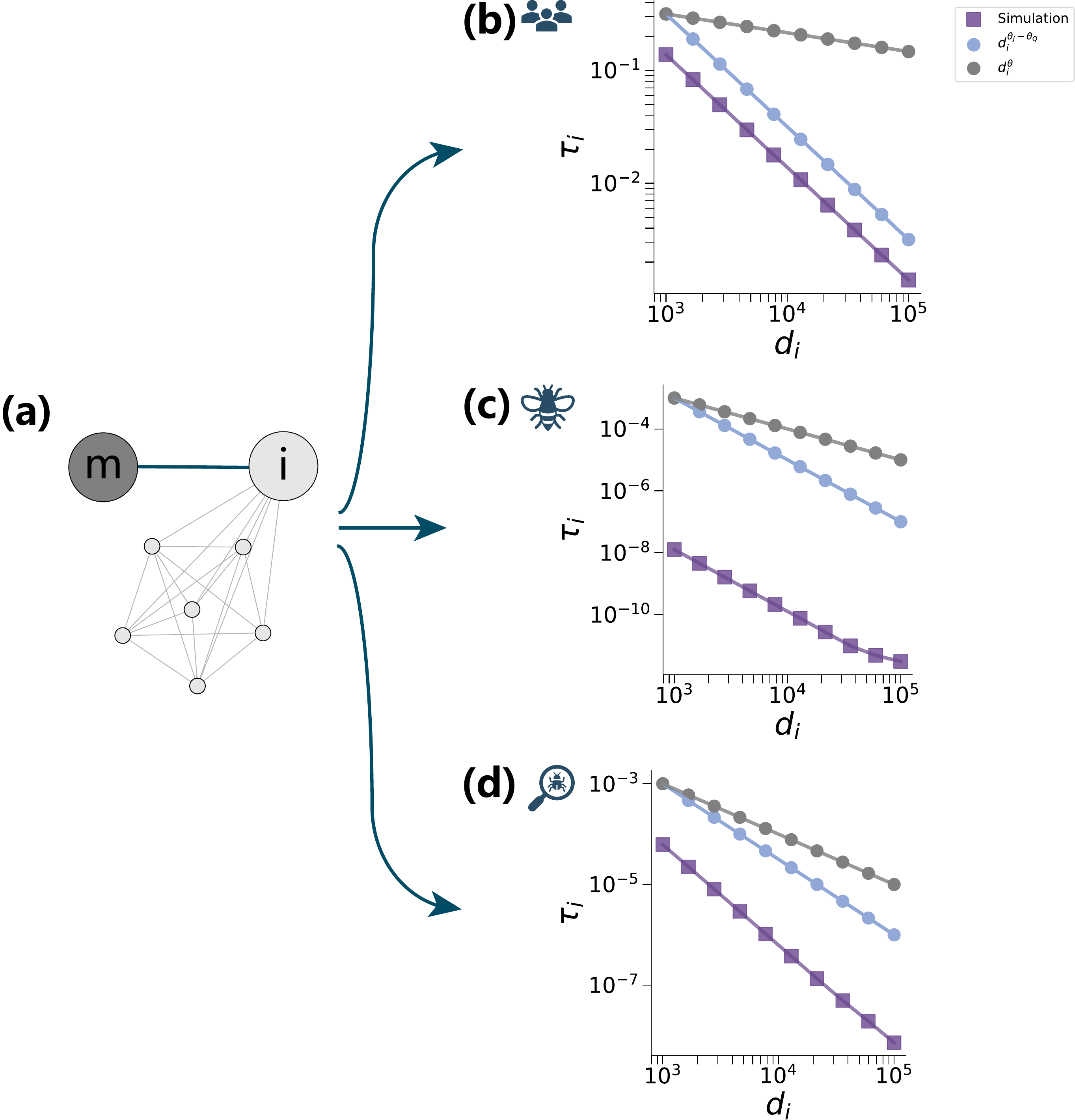}
\caption{
\label{fig:3}
\textbf{Local propagation time for cliques based on independent edges.}
\textbf{(a)} Network with perturbation at $m$ and target to node i having $d_i$ edges, each node connecting to $i$ connects to each other.
\textbf{(b)} Propagation time $\tau_i$ simulated on network (a) and population dynamics proposed in Eq.(\ref{equ:dynamic:P}): $\dot{x}_i(t)=-Bx_i^a(t)+\sum\limits_{j=1}^NA_{ij}x_j^b(t)$,  with $B=0.1, a=1.2, b=1, \alpha=0.01(\theta_J, \theta_Q<0)$. Comparison of scaling coefficient Eqs.(\ref{equ:hens:1})-(\ref{equ:ours:1}) and simulation. The trend of simulation propagation time agrees to our prediction $\tau_i\sim d_i^{\theta_J-\theta_Q}$ in Eq.(\ref{equ:ours:1}), but disagrees the prediction $\tau_i\sim d_i^{\theta_J}$(gray) in Eq.(\ref{equ:hens:1}).
\textbf{(c)} Simulation on network (a) and Mutualistic dynamics proposed in Eq.(\ref{equ:dynamic:M}): $\dot{x}_i(t)=-Bx_i(t)\left(1-\frac{x_i(t)}{C}\right)+x_i(t)\sum\limits_{j=1}^NA_{ij}x_j(t)$,  with $B=0.1, C=2, \alpha=0.01(\theta_J, \theta_Q<0)$.
\textbf{(d)} Simulation on network (a) and inhibitory dynamics proposed in Eq.(\ref{equ:dynamic:I}): $\dot{x}_i(t)=-Bx_i(t)\left(1-\frac{x_i(t)}{C}\right)^2+x_i(t)\sum\limits_{j=1}^NA_{ij}x_j(t)$,  with $B=0.1, C=2, \alpha=0.01(\theta_J, \theta_Q<0)$.
}
\end{figure*}
\par In the previous discussion, we
analyzed cliques based on triangles. However, in real world, some perturbation can occur in ordinary nodes, which means the case of clique except source nodes is also important. Following our theoretical framework, the perturbed jacobian matrix $\tilde{\mathbf{J}}$ is
\begin{equation}
    \begin{aligned}
        \label{equ:result:cliqueind:0}
        \tilde{\mathbf{J}}=
        \begin{bmatrix}
        0 & 0 & 0 & \cdots & 0 & 0\\
        0 & -b_2 & a_{23} &\cdots & a_{2,t-1} & a_{2t}\\
        0 & a_{32} & b_3 &\cdots & a_{3,t-11} & a_{3t}\\
        \vdots & \vdots & \vdots & \ddots & \vdots & \vdots\\
        0 & a_{t-1,2} & a_{t-1,3} & \cdots & -b_{t-1} & -a_{t-1,t}\\
        a_{t1} & a_{t2} & a_{t3} & \cdots & a_{t,t-1} & -b_{t}
        \end{bmatrix}
    \end{aligned},
\end{equation}
in which $t=s_{im}+2$ is the index for node $i$, $i_1,i_2,i_{t-1}$ are indexs for independent edges of node $i$, and
\begin{equation*}
    \begin{aligned}
        \left\{
        \begin{array}{l}
            a_{t1}=H_1(x_i^*)A_{im}H_2^{'}(x_m^*),\\
            a_{tk}=H_1(x_i^*)A_{ii_k}H_2^{'}(x_{i_k}^*),\\
            a_{kt}=H_1(x_{i_k}^*)A_{i_ki}H_2^{'}(x_i^*),\\
            b_k=-H_1(x_{i_k}^*)\left[\frac{F(x_{i_k}^*)}{H_1(x_{i_k}^*)}\right]^{'},\\
            b_t=-H_1(x_{i}^*)\left[\frac{F(x_{i}^*)}{H_1(x_{i}^*)}\right]^{'}.
        \end{array}
        \right.
        ,k=2,t_{im}+1.
    \end{aligned}
\end{equation*}
Through definition of $\tilde{\mathbf{J}}$ in Eq.(\ref{equ:result:cliqueind:0}), we calculate items $c_{ij}$ for in $(sI-\tilde{\mathbf{J}})^{-1}$ and $c_{ij}$ satisfies following linear system:
\begin{equation}
    \begin{aligned}
        \label{equ:result:cliqueind:1}
        \begin{bmatrix}
        0 & s+b_2 &\cdots & -a_{2,t-1} & -a_{2t}\\
        0 & -a_{32} &\cdots & -a_{3,t-11} & -a_{3t}\\
        \vdots & \vdots & \ddots & \vdots & \vdots\\
        0 & -a_{t-1,2} & \cdots & s+b_{t-1} & -a_{t-1,t}\\
        -a_{t1} & -a_{t2} & \cdots & -a_{t,t-1} & s+b_{t}
        \end{bmatrix}
        \begin{bmatrix}
        \frac{1}{s} \\
        c_{21} \\
        \vdots \\
        c_{t-1,1} \\
        c_{t1}
        \end{bmatrix}=0.
    \end{aligned}
\end{equation}
This linear system Eq.(\ref{equ:result:cliqueind:1}) can be simplified after multiplying the following matrix to the left
\begin{equation*}
    \begin{aligned}
        \begin{bmatrix}
            1 & 0 &\cdots & 0 & -f_2\\
            0 & 1 &\cdots & 0 & -f_3\\
            \vdots & \vdots & \ddots & \vdots & \vdots\\
            0 & 0 & \cdots & 1 & -f_{t-1}\\
            0 & 0 & \cdots & 0 & 1
        \end{bmatrix},
    \end{aligned}
\end{equation*}
in which $f_k=-\frac{H_1(x_{i_k})}{H_1(x_i^*)},k=2,t_{im}+1$, we obtain
\begin{equation}
    \begin{aligned}
        \label{equ:result:cliqueind:2}
        \begin{bmatrix}
        f_2a_{t1} & s+b_2+f_2a_{t2} &\cdots & -a_{2t}-f_2(s+b_t)\\
        f_3a_{t1} & 0 &\cdots & -a_{3t}-f_3(s+b_t)\\
        \vdots & \vdots & \ddots & \vdots\\
        f_{t-1}a_{t1} & 0& \cdots & -a_{t-1,t}-f_{t-1}(s+b_t)\\
        -a_{t1} & -a_{t2} & \cdots & s+b_{t}
        \end{bmatrix}
        \begin{bmatrix}
        \frac{1}{s} \\
        c_{21} \\
        \vdots \\
        c_{t1}
        \end{bmatrix}=0.
    \end{aligned}
\end{equation}
Preforming linear transformations for Eq.(\ref{equ:result:cliqueind:2}), we can obtain value for $c_{21}$, which captures value of $\Delta x_i(t)$:
\begin{equation}
    \begin{aligned}
        \label{equ:result:cliqueind:3}
        c_{t1}&=\frac{\left(a_{t1}-\sum\limits_{k=2}^{s_{im}+1}a_{tk}f_ka_{t1}\right)\frac{1}{s}}{s+b_t-\sum\limits_{k=2}^{s_{im}+1}\frac{a_{tk}\left(a_{kt}+f_k(s+b_t)\right)}{s+b_k+f_ka_{tk}}}\\
        &\approx \frac{\frac{\tilde{a}_{t1}}{1-e_1}}{s+b_t-\frac{e_2}{1-e_1}},
    \end{aligned}
\end{equation}
where we ignore $s$ in the denominator under the condition $b_k+f_ka_{tk}$ is large, i.e. $\theta_J<0$, and
\begin{equation*}
    \begin{aligned}
        \tilde{a}_{t1}&=a_{t1}-\sum\limits_{k=2}^{s_{im}+1}a_{tk}f_ka_{t1}\\
        &=H_1(x_i^*)A_{im}H_2^{'}(x_m^*)\left(1-\sum\limits_{k=2}^{s_{im}+1}A_{ii_k}H_2^{'}(x_{i_k}^*)H_1(x_{i_k}^*)\right)\\
        &\approx H_1(x_i^*)A_{im}H_2^{'}(x_m^*)=a_{t1}.
    \end{aligned}
\end{equation*}
The value of $\Delta x_i(t)$ can be approximately obtained through inverse laplacian transformation for Eq.(\ref{equ:result:cliqueind:3}), and
\begin{equation}
    \begin{aligned}
        \label{equ:result:cliqueind:4}
        \Delta x_i(t)&=\mathcal{L}^{-1}(c_{t1})\Delta x_m\\
        &=\frac{\frac{\tilde{a}_{t1}}{1-e_1}\left(1-\exp\left(-\left(b_t-\frac{e_2}{1-e_1}\right)t\right)\right)}{b_t-\frac{e_2}{1-e_1}}\Delta x_m.
    \end{aligned}
\end{equation}
\par Shifted state for node $i$'s message is the value of Eq.(\ref{equ:result:cliqueind:4}) when $t\to\infty$ and
\begin{equation}
    \begin{aligned}
        \Delta x_i(\infty)&=\frac{a_{t1}}{b_t-e_2}\Delta x_m\\
        &=\frac{A_{im}H_2^{'}(x_m^*)\Delta x_m}{-\left[\frac{F(x_{i}^*)}{H_1(x_{i}^*)}\right]^{'}+H_2^{'}(x_i^*)\sum\limits_{k=2}^{s_{im}+1}\frac{A_{ii_k}A_{i_ki}+A_{ii_k}C^{-1}(x_{i}^*)}{C^{-1}(x_{i_k}^*)+A_{ii_k}}},
    \end{aligned}
\end{equation}
while propagation time can be obtained by $\eta$-scale for Eq.(\ref{equ:result:cliqueind:4}), i.e.
\begin{equation}
    \begin{aligned}
        \eta&=1-\exp\left(-\left(b_t-\frac{e_2}{1-e_1}\right)\tau_i\right),
    \end{aligned}
\end{equation}
and
\begin{equation}
    \begin{aligned}
        \label{equ:result:cliqueind:5}
        \tau_i&=-\frac{\ln(1-\eta)}{\frac{b_t-e_2}{1-e_1}}\\
        &=\frac{\ln(1-\eta)}{H_1(x_{i}^*)\left[\frac{F(x_{i}^*)}{H_1(x_{i}^*)}\right]^{'}}\frac{1-\sum\limits_{k=2}^{s_{im}+1}\frac{A_{ii_k}}{C^{-1}(x_{i_k}^*)+A_{ii_k}}}{1+C(x_i^*)\sum\limits_{k=2}^{s_{im}+1}\frac{A_{ii_k}A_{i_ki}+A_{ii_k}C^{-1}(x_{i}^*)}{C^{-1}(x_{i_k}^*)+A_{ii_k}}}.
    \end{aligned}
\end{equation}
For $\theta_J<0$ and $\theta_Q<0$ propagation time Eq.(\ref{equ:result:cliqueind:5}) can be simplified as
\begin{equation}
    \begin{aligned}
        \tau_i&\sim d_i^{\theta_J},
    \end{aligned}
\end{equation}
and for $\theta_J<0$ and $\theta_Q>0$ propagation time Eq.(\ref{equ:result:cliqueind:5}) can be simplified as
\begin{equation}
    \begin{aligned}
        \tau_i&\sim d_i^{\theta_J-\theta_Q}.
    \end{aligned}
\end{equation}

\par\
\par We have
proposed
propagation analysis
for the above
basic motifs.
However,
in real
world,
these
ideal situations
may not be
achieved.
Based on
the decomposition of network
introduced in Fig.\ref{fig:1},
we can decompose
a network
into
combination of
four
basic motifs,
i.e.
triangles,
independent edges,
cliques based on triangles
and
cliques based on independent edges.
Therefore,
the actual propagation time
for one node in
the network
can be
expressed as
combination of
propagation time
for these four
motifs.
Since
proportion of
each motif is
uncertain,
we can provide
rigorous criteria
for
propagation time
at
each node,
considering
its global topology,
this is
useful and essential
for real network analysis
and dynamic analysis.

\section{Discussion and Outlook}
\par
Network structures,
such as
triangles
and cliques,
are always
the
focal point of
research
in
complex networks.
In this
paper,
we
introduce
a powerful
and
analytical framework to
decouple and
examine
signal propagation
within
complex
network motifs.
We
present
specific expressions
for
propagation time
under
various
basic network motifs
and dynamic models.
Our
conclusions
indicate
the substantial impact of
dense network structures
on
signal propagation,
since
sufficient interactions between nodes
profoundly
influence
the propagation mode of
the system.
This phenomenon
is not solely a consequence of
the local topology of nodes,
but
is primarily
driven by the
generation of a large number of cycles.
These cycles
can
lead to
large differences in the propagation time,
often spanning
orders of magnitude,
as observed in phenomena
such as
population flow
between cities.
Distinguishing itself from
the
undirected propagation pattern
proposed in
previous framework,
our
framework
reveals the mathematical principle
underlying
the
negative feedback mechanism.
Leveraging
our
powerful
decoupling
methods,
we can
isolate
the mutual
interactions
between nodes.
This ability
enhances
our understanding
on
this
ubiquitous mechanism in real
world
and
augments our capacity for
preventing and managing
catastrophic events,
such as species extinction.

\par
Furthermore,
our framework
liberates us
from
the constraints of
fixed
perturbation
and
the absence of
time-delay
effects
in
signal propagation framework.
We can
now
set state of the source node
based on
a fixed differential equation,
or introduce time-delay
to neighbourhood nodes,
which
significantly
broadens
the
applicability of signal propagation framework
and
aligns them more effectively with
specific physical phenomena in
the
real world.
Additionally,
our
decoupling
method
is adaptable for
analyzing
the
signal propagation process
in
high-dimensional networks
and
high-order dynamics.
However,
it's important to note that
our framework
is still
constrained by
conditions
involving
small perturbations.
Small perturbations can
disturb
the state of nodes 
near their stable state,
enabling the application of
linear theory.
Yet,
for
the analysis
of
large perturbations,
there exists
an unmet needs
as analysis tools
for such
cases
remain
unsolved\cite{Bianconi2023ComplexSI},
this problem remains
to be
addressed
in future research.
\par\
\par The data and code for article and Supplement Information will be publicly available upon publication.
\par Partly supported by the National Natural Science Foundation of China (Nos.12371354, 11971311, 12161141003) and Science and Technology Commission of Shanghai Municipality, China (No. 22JC1403602), National Key R\&D Program of China under Grant No. 2022YFA1006400 and the Fundamental Research Funds for the Central Universities, China.


\begin{thebibliography}{31}
\expandafter\ifx\csname natexlab\endcsname\relax\def\natexlab#1{#1}\fi
\expandafter\ifx\csname bibnamefont\endcsname\relax
  \def\bibnamefont#1{#1}\fi
\expandafter\ifx\csname bibfnamefont\endcsname\relax
  \def\bibfnamefont#1{#1}\fi
\expandafter\ifx\csname citenamefont\endcsname\relax
  \def\citenamefont#1{#1}\fi
\expandafter\ifx\csname url\endcsname\relax
  \def\url#1{\texttt{#1}}\fi
\expandafter\ifx\csname urlprefix\endcsname\relax\def\urlprefix{URL }\fi
\providecommand{\bibinfo}[2]{#2}
\providecommand{\eprint}[2][]{\url{#2}}

\bibitem[{\citenamefont{Balcan et~al.}(2009)\citenamefont{Balcan, Colizza,
  Gonçalves, Hu, Ramasco, and Vespignani}}]{Balcan2009MultiscaleMN}
\bibinfo{author}{\bibfnamefont{D.}~\bibnamefont{Balcan}},
  \bibinfo{author}{\bibfnamefont{V.}~\bibnamefont{Colizza}},
  \bibinfo{author}{\bibfnamefont{B.}~\bibnamefont{Gonçalves}},
  \bibinfo{author}{\bibfnamefont{H.}~\bibnamefont{Hu}},
  \bibinfo{author}{\bibfnamefont{J.~J.} \bibnamefont{Ramasco}},
  \bibnamefont{and}
  \bibinfo{author}{\bibfnamefont{A.}~\bibnamefont{Vespignani}},
  \bibinfo{journal}{Proceedings of the National Academy of Sciences}
  \textbf{\bibinfo{volume}{106}}, \bibinfo{pages}{21484 }
  (\bibinfo{year}{2009}).

\bibitem[{\citenamefont{Bornholdt}(2008)}]{Bornholdt2008BooleanNM}
\bibinfo{author}{\bibfnamefont{S.}~\bibnamefont{Bornholdt}},
  \bibinfo{journal}{Journal of The Royal Society Interface}
  \textbf{\bibinfo{volume}{5}}, \bibinfo{pages}{S85} (\bibinfo{year}{2008}).

\bibitem[{\citenamefont{Balaji et~al.}(2006)\citenamefont{Balaji, Babu, Iyer,
  Luscombe, and Aravind}}]{Balaji2006ComprehensiveAO}
\bibinfo{author}{\bibfnamefont{S.}~\bibnamefont{Balaji}},
  \bibinfo{author}{\bibfnamefont{M.~M.} \bibnamefont{Babu}},
  \bibinfo{author}{\bibfnamefont{L.~M.} \bibnamefont{Iyer}},
  \bibinfo{author}{\bibfnamefont{N.~M.} \bibnamefont{Luscombe}},
  \bibnamefont{and} \bibinfo{author}{\bibfnamefont{L.}~\bibnamefont{Aravind}},
  \bibinfo{journal}{Journal of Molecular Biology}
  \textbf{\bibinfo{volume}{360}}, \bibinfo{pages}{213} (\bibinfo{year}{2006}).

\bibitem[{\citenamefont{Rand et~al.}(2021)\citenamefont{Rand, Raju, S{\'a}ez,
  Corson, and Siggia}}]{Rand2021GeometryOG}
\bibinfo{author}{\bibfnamefont{D.~A.} \bibnamefont{Rand}},
  \bibinfo{author}{\bibfnamefont{A.}~\bibnamefont{Raju}},
  \bibinfo{author}{\bibfnamefont{M.}~\bibnamefont{S{\'a}ez}},
  \bibinfo{author}{\bibfnamefont{F.}~\bibnamefont{Corson}}, \bibnamefont{and}
  \bibinfo{author}{\bibfnamefont{E.~D.} \bibnamefont{Siggia}},
  \bibinfo{journal}{Proceedings of the National Academy of Sciences}
  \textbf{\bibinfo{volume}{118}}, \bibinfo{pages}{e2109729118}
  (\bibinfo{year}{2021}).

\bibitem[{\citenamefont{Karlebach and Shamir}(2008)}]{Karlebach2008ModellingAA}
\bibinfo{author}{\bibfnamefont{G.}~\bibnamefont{Karlebach}} \bibnamefont{and}
  \bibinfo{author}{\bibfnamefont{R.}~\bibnamefont{Shamir}},
  \bibinfo{journal}{Nature Reviews Molecular Cell Biology}
  \textbf{\bibinfo{volume}{9}}, \bibinfo{pages}{770} (\bibinfo{year}{2008}).

\bibitem[{\citenamefont{Li and Wang}(2014)}]{Li2014LandscapeAF}
\bibinfo{author}{\bibfnamefont{C.}~\bibnamefont{Li}} \bibnamefont{and}
  \bibinfo{author}{\bibfnamefont{J.}~\bibnamefont{Wang}},
  \bibinfo{journal}{Proceedings of the National Academy of Sciences}
  \textbf{\bibinfo{volume}{111}}, \bibinfo{pages}{14130 }
  (\bibinfo{year}{2014}).

\bibitem[{\citenamefont{Kumar et~al.}(2010)\citenamefont{Kumar, Rotter, and
  Aertsen}}]{Kumar2010SpikingAP}
\bibinfo{author}{\bibfnamefont{A.}~\bibnamefont{Kumar}},
  \bibinfo{author}{\bibfnamefont{S.}~\bibnamefont{Rotter}}, \bibnamefont{and}
  \bibinfo{author}{\bibfnamefont{A.}~\bibnamefont{Aertsen}},
  \bibinfo{journal}{Nature Reviews Neuroscience} \textbf{\bibinfo{volume}{11}},
  \bibinfo{pages}{615} (\bibinfo{year}{2010}).

\bibitem[{\citenamefont{Hens et~al.}(2019)\citenamefont{Hens, Harush, Haber,
  Cohen, and Barzel}}]{Hens2019SpatiotemporalSP}
\bibinfo{author}{\bibfnamefont{C.}~\bibnamefont{Hens}},
  \bibinfo{author}{\bibfnamefont{U.}~\bibnamefont{Harush}},
  \bibinfo{author}{\bibfnamefont{S.}~\bibnamefont{Haber}},
  \bibinfo{author}{\bibfnamefont{R.}~\bibnamefont{Cohen}}, \bibnamefont{and}
  \bibinfo{author}{\bibfnamefont{B.}~\bibnamefont{Barzel}},
  \bibinfo{journal}{Nature Physics} \textbf{\bibinfo{volume}{15}},
  \bibinfo{pages}{403} (\bibinfo{year}{2019}).

\bibitem[{\citenamefont{Milo et~al.}(2002)\citenamefont{Milo, Shen-Orr,
  Itzkovitz, Kashtan, Chklovskii, and Alon}}]{Milo2002NetworkMS}
\bibinfo{author}{\bibfnamefont{R.}~\bibnamefont{Milo}},
  \bibinfo{author}{\bibfnamefont{S.~S.} \bibnamefont{Shen-Orr}},
  \bibinfo{author}{\bibfnamefont{S.}~\bibnamefont{Itzkovitz}},
  \bibinfo{author}{\bibfnamefont{N.}~\bibnamefont{Kashtan}},
  \bibinfo{author}{\bibfnamefont{D.~B.} \bibnamefont{Chklovskii}},
  \bibnamefont{and} \bibinfo{author}{\bibfnamefont{U.}~\bibnamefont{Alon}},
  \bibinfo{journal}{Science} \textbf{\bibinfo{volume}{298}},
  \bibinfo{pages}{824} (\bibinfo{year}{2002}).

\bibitem[{\citenamefont{Menck et~al.}(2013)\citenamefont{Menck, Heitzig,
  Marwan, and Kurths}}]{Menck2013HowBS}
\bibinfo{author}{\bibfnamefont{P.~J.} \bibnamefont{Menck}},
  \bibinfo{author}{\bibfnamefont{J.}~\bibnamefont{Heitzig}},
  \bibinfo{author}{\bibfnamefont{N.}~\bibnamefont{Marwan}}, \bibnamefont{and}
  \bibinfo{author}{\bibfnamefont{J.}~\bibnamefont{Kurths}},
  \bibinfo{journal}{Nature Physics} \textbf{\bibinfo{volume}{9}},
  \bibinfo{pages}{89 } (\bibinfo{year}{2013}).

\bibitem[{\citenamefont{Menck et~al.}(2014)\citenamefont{Menck, Heitzig,
  Kurths, and Schellnhuber}}]{Menck2014HowDE}
\bibinfo{author}{\bibfnamefont{P.~J.} \bibnamefont{Menck}},
  \bibinfo{author}{\bibfnamefont{J.}~\bibnamefont{Heitzig}},
  \bibinfo{author}{\bibfnamefont{J.}~\bibnamefont{Kurths}}, \bibnamefont{and}
  \bibinfo{author}{\bibfnamefont{H.~J.} \bibnamefont{Schellnhuber}},
  \bibinfo{journal}{Nature Communications} \textbf{\bibinfo{volume}{5}},
  \bibinfo{pages}{3969} (\bibinfo{year}{2014}).

\bibitem[{\citenamefont{Girvan and Newman}(2001)}]{Girvan2001CommunitySI}
\bibinfo{author}{\bibfnamefont{M.}~\bibnamefont{Girvan}} \bibnamefont{and}
  \bibinfo{author}{\bibfnamefont{M.~E.~J.} \bibnamefont{Newman}},
  \bibinfo{journal}{Proceedings of the National Academy of Sciences}
  \textbf{\bibinfo{volume}{99}}, \bibinfo{pages}{7821 } (\bibinfo{year}{2001}).

\bibitem[{\citenamefont{Alon}(2007)}]{Alon2007NetworkMT}
\bibinfo{author}{\bibfnamefont{U.}~\bibnamefont{Alon}},
  \bibinfo{journal}{Nature Reviews Genetics} \textbf{\bibinfo{volume}{8}},
  \bibinfo{pages}{450} (\bibinfo{year}{2007}).

\bibitem[{\citenamefont{Lambiotte et~al.}(2019)\citenamefont{Lambiotte,
  Rosvall, and Scholtes}}]{Lambiotte2019FromNT}
\bibinfo{author}{\bibfnamefont{R.}~\bibnamefont{Lambiotte}},
  \bibinfo{author}{\bibfnamefont{M.}~\bibnamefont{Rosvall}}, \bibnamefont{and}
  \bibinfo{author}{\bibfnamefont{I.}~\bibnamefont{Scholtes}},
  \bibinfo{journal}{Nature Physics} \textbf{\bibinfo{volume}{15}},
  \bibinfo{pages}{313 } (\bibinfo{year}{2019}).

\bibitem[{\citenamefont{Battiston et~al.}(2021)\citenamefont{Battiston, Amico,
  Barrat, Bianconi, de~Arruda, Franceschiello, Iacopini, K{\'e}fi, Latora,
  Moreno et~al.}}]{Battiston2021ThePO}
\bibinfo{author}{\bibfnamefont{F.}~\bibnamefont{Battiston}},
  \bibinfo{author}{\bibfnamefont{E.}~\bibnamefont{Amico}},
  \bibinfo{author}{\bibfnamefont{A.}~\bibnamefont{Barrat}},
  \bibinfo{author}{\bibfnamefont{G.}~\bibnamefont{Bianconi}},
  \bibinfo{author}{\bibfnamefont{G.~F.} \bibnamefont{de~Arruda}},
  \bibinfo{author}{\bibfnamefont{B.}~\bibnamefont{Franceschiello}},
  \bibinfo{author}{\bibfnamefont{I.}~\bibnamefont{Iacopini}},
  \bibinfo{author}{\bibfnamefont{S.}~\bibnamefont{K{\'e}fi}},
  \bibinfo{author}{\bibfnamefont{V.}~\bibnamefont{Latora}},
  \bibinfo{author}{\bibfnamefont{Y.}~\bibnamefont{Moreno}},
  \bibnamefont{et~al.}, \bibinfo{journal}{Nature Physics}
  \textbf{\bibinfo{volume}{17}}, \bibinfo{pages}{1093 } (\bibinfo{year}{2021}).

\bibitem[{\citenamefont{Battiston et~al.}(2020)\citenamefont{Battiston,
  Cencetti, Iacopini, Latora, Lucas, Patania, Young, and
  Petri}}]{Battiston2020NetworksBP}
\bibinfo{author}{\bibfnamefont{F.}~\bibnamefont{Battiston}},
  \bibinfo{author}{\bibfnamefont{G.}~\bibnamefont{Cencetti}},
  \bibinfo{author}{\bibfnamefont{I.}~\bibnamefont{Iacopini}},
  \bibinfo{author}{\bibfnamefont{V.}~\bibnamefont{Latora}},
  \bibinfo{author}{\bibfnamefont{M.}~\bibnamefont{Lucas}},
  \bibinfo{author}{\bibfnamefont{A.}~\bibnamefont{Patania}},
  \bibinfo{author}{\bibfnamefont{J.-G.} \bibnamefont{Young}}, \bibnamefont{and}
  \bibinfo{author}{\bibfnamefont{G.}~\bibnamefont{Petri}},
  \bibinfo{journal}{Physics Reports} \textbf{\bibinfo{volume}{874}},
  \bibinfo{pages}{1} (\bibinfo{year}{2020}).

\bibitem[{\citenamefont{St-Onge et~al.}(2021)\citenamefont{St-Onge, Sun,
  Allard, H'ebert-Dufresne, and Bianconi}}]{StOnge2021UniversalNI}
\bibinfo{author}{\bibfnamefont{G.}~\bibnamefont{St-Onge}},
  \bibinfo{author}{\bibfnamefont{H.}~\bibnamefont{Sun}},
  \bibinfo{author}{\bibfnamefont{A.}~\bibnamefont{Allard}},
  \bibinfo{author}{\bibfnamefont{L.}~\bibnamefont{H'ebert-Dufresne}},
  \bibnamefont{and} \bibinfo{author}{\bibfnamefont{G.}~\bibnamefont{Bianconi}},
  \bibinfo{journal}{Physical Review Letters} \textbf{\bibinfo{volume}{127}},
  \bibinfo{pages}{158301} (\bibinfo{year}{2021}).

\bibitem[{\citenamefont{Bao et~al.}(2022)\citenamefont{Bao, Hu, Ji, Lin,
  Kurths, and Nagler}}]{Bao2022ImpactOB}
\bibinfo{author}{\bibfnamefont{X.}~\bibnamefont{Bao}},
  \bibinfo{author}{\bibfnamefont{Q.}~\bibnamefont{Hu}},
  \bibinfo{author}{\bibfnamefont{P.}~\bibnamefont{Ji}},
  \bibinfo{author}{\bibfnamefont{W.}~\bibnamefont{Lin}},
  \bibinfo{author}{\bibfnamefont{J.}~\bibnamefont{Kurths}}, \bibnamefont{and}
  \bibinfo{author}{\bibfnamefont{J.}~\bibnamefont{Nagler}},
  \bibinfo{journal}{Nature Communications} \textbf{\bibinfo{volume}{13}},
  \bibinfo{pages}{5301} (\bibinfo{year}{2022}).

\bibitem[{\citenamefont{Castellano et~al.}(2009)\citenamefont{Castellano,
  Fortunato, and Loreto}}]{Castellano2007StatisticalPO}
\bibinfo{author}{\bibfnamefont{C.}~\bibnamefont{Castellano}},
  \bibinfo{author}{\bibfnamefont{S.}~\bibnamefont{Fortunato}},
  \bibnamefont{and} \bibinfo{author}{\bibfnamefont{V.}~\bibnamefont{Loreto}},
  \bibinfo{journal}{Reviews of Modern Physics} \textbf{\bibinfo{volume}{81}},
  \bibinfo{pages}{591} (\bibinfo{year}{2009}).

\bibitem[{\citenamefont{Dodds and Watts}(2005)}]{Dodds2005AGM}
\bibinfo{author}{\bibfnamefont{P.}~\bibnamefont{Dodds}} \bibnamefont{and}
  \bibinfo{author}{\bibfnamefont{D.}~\bibnamefont{Watts}},
  \bibinfo{journal}{Journal of Theoretical Biology}
  \textbf{\bibinfo{volume}{232}}, \bibinfo{pages}{587} (\bibinfo{year}{2005}).

\bibitem[{\citenamefont{May}(1976)}]{May1976SimpleMM}
\bibinfo{author}{\bibfnamefont{R.~M.} \bibnamefont{May}},
  \bibinfo{journal}{Nature} \textbf{\bibinfo{volume}{261}},
  \bibinfo{pages}{459} (\bibinfo{year}{1976}).

\bibitem[{\citenamefont{Voit}(2000)}]{Voit2000ComputationalAO}
\bibinfo{author}{\bibfnamefont{E.~O.} \bibnamefont{Voit}},
  \emph{\bibinfo{title}{Computational Analysis of Biochemical Systems: A
  Practical Guide for Biochemists and Molecular Biologists}}
  (\bibinfo{publisher}{Cambridge University Press}, \bibinfo{year}{2000}).

\bibitem[{\citenamefont{Harush and Barzel}(2017)}]{Harush2017DynamicPO}
\bibinfo{author}{\bibfnamefont{U.}~\bibnamefont{Harush}} \bibnamefont{and}
  \bibinfo{author}{\bibfnamefont{B.}~\bibnamefont{Barzel}},
  \bibinfo{journal}{Nature Communications} \textbf{\bibinfo{volume}{8}},
  \bibinfo{pages}{2181} (\bibinfo{year}{2017}).

\bibitem[{\citenamefont{Meena et~al.}(2023)\citenamefont{Meena, Hens, Acharyya,
  Haber, Boccaletti, and Barzel}}]{Meena2020EmergentSI}
\bibinfo{author}{\bibfnamefont{C.}~\bibnamefont{Meena}},
  \bibinfo{author}{\bibfnamefont{C.}~\bibnamefont{Hens}},
  \bibinfo{author}{\bibfnamefont{S.}~\bibnamefont{Acharyya}},
  \bibinfo{author}{\bibfnamefont{S.}~\bibnamefont{Haber}},
  \bibinfo{author}{\bibfnamefont{S.}~\bibnamefont{Boccaletti}},
  \bibnamefont{and} \bibinfo{author}{\bibfnamefont{B.}~\bibnamefont{Barzel}},
  \bibinfo{journal}{Nature Physics} \textbf{\bibinfo{volume}{19}},
  \bibinfo{pages}{1033–1042} (\bibinfo{year}{2023}).

\bibitem[{\citenamefont{Alon}(2019)}]{Alon2019AnIT}
\bibinfo{author}{\bibfnamefont{U.}~\bibnamefont{Alon}},
  \emph{\bibinfo{title}{An introduction to systems biology : design principles
  of biological circuits}} (\bibinfo{publisher}{Chapman and Hall/CRC},
  \bibinfo{year}{2019}).

\bibitem[{\citenamefont{Barzel and Biham}(2011)}]{Barzel2011BinomialME}
\bibinfo{author}{\bibfnamefont{B.}~\bibnamefont{Barzel}} \bibnamefont{and}
  \bibinfo{author}{\bibfnamefont{O.}~\bibnamefont{Biham}},
  \bibinfo{journal}{Physical Review Letters} \textbf{\bibinfo{volume}{106}},
  \bibinfo{pages}{150602} (\bibinfo{year}{2011}).

\bibitem[{\citenamefont{Novozhilov et~al.}(2006)\citenamefont{Novozhilov,
  Karev, and Koonin}}]{Novozhilov2006BiologicalAO}
\bibinfo{author}{\bibfnamefont{A.~S.} \bibnamefont{Novozhilov}},
  \bibinfo{author}{\bibfnamefont{G.~P.} \bibnamefont{Karev}}, \bibnamefont{and}
  \bibinfo{author}{\bibfnamefont{E.~V.} \bibnamefont{Koonin}},
  \bibinfo{journal}{Briefings in Bioinformatics} \textbf{\bibinfo{volume}{7}},
  \bibinfo{pages}{70} (\bibinfo{year}{2006}).

\bibitem[{\citenamefont{Hayes and Babu}(2004)}]{Hayes2004ModelingAA}
\bibinfo{author}{\bibfnamefont{J.~F.} \bibnamefont{Hayes}} \bibnamefont{and}
  \bibinfo{author}{\bibfnamefont{T.~V. J.~G.} \bibnamefont{Babu}},
  \emph{\bibinfo{title}{Modeling and Analysis of Telecommunications Networks}}
  (\bibinfo{publisher}{John Wiley \& Sons, Ltd}, \bibinfo{year}{2004}).

\bibitem[{\citenamefont{Holling}(1959)}]{Holling1959SomeCO}
\bibinfo{author}{\bibfnamefont{C.~S.} \bibnamefont{Holling}},
  \bibinfo{journal}{The Canadian Entomologist} \textbf{\bibinfo{volume}{91}},
  \bibinfo{pages}{385} (\bibinfo{year}{1959}).

\bibitem[{\citenamefont{Barzel et~al.}(2015)\citenamefont{Barzel, Liu, and
  Barab{\'a}si}}]{Barzel2015ConstructingMM}
\bibinfo{author}{\bibfnamefont{B.}~\bibnamefont{Barzel}},
  \bibinfo{author}{\bibfnamefont{Y.-Y.} \bibnamefont{Liu}}, \bibnamefont{and}
  \bibinfo{author}{\bibfnamefont{A.-L.} \bibnamefont{Barab{\'a}si}},
  \bibinfo{journal}{Nature Communications} \textbf{\bibinfo{volume}{6}},
  \bibinfo{pages}{7186} (\bibinfo{year}{2015}).

\bibitem[{\citenamefont{Bianconi et~al.}(2023)\citenamefont{Bianconi, Arenas,
  Biamonte, Carr, Kahng, Kert{\'e}sz, Kurths, L{\"u}, Masoller, Motter
  et~al.}}]{Bianconi2023ComplexSI}
\bibinfo{author}{\bibfnamefont{G.}~\bibnamefont{Bianconi}},
  \bibinfo{author}{\bibfnamefont{A.}~\bibnamefont{Arenas}},
  \bibinfo{author}{\bibfnamefont{J.~D.} \bibnamefont{Biamonte}},
  \bibinfo{author}{\bibfnamefont{L.~D.} \bibnamefont{Carr}},
  \bibinfo{author}{\bibfnamefont{B.}~\bibnamefont{Kahng}},
  \bibinfo{author}{\bibfnamefont{J.}~\bibnamefont{Kert{\'e}sz}},
  \bibinfo{author}{\bibfnamefont{J.}~\bibnamefont{Kurths}},
  \bibinfo{author}{\bibfnamefont{L.}~\bibnamefont{L{\"u}}},
  \bibinfo{author}{\bibfnamefont{C.}~\bibnamefont{Masoller}},
  \bibinfo{author}{\bibfnamefont{A.~E.} \bibnamefont{Motter}},
  \bibnamefont{et~al.}, \bibinfo{journal}{Journal of Physics: Complexity}
  \textbf{\bibinfo{volume}{4}}, \bibinfo{pages}{010201} (\bibinfo{year}{2023}).

\end{thebibliography}
\end{document}